\documentclass[aps,prl,twocolumn, superscriptaddress, nofootinbib,10pt]{revtex4-2}
\usepackage[utf8]{inputenc}
\usepackage{graphicx,epsfig,overpic}
\usepackage{amsmath,amsfonts,amssymb,mathtools}
\usepackage{bm,physics,enumitem}
\usepackage[dvipsnames]{xcolor}
\usepackage[colorlinks,citecolor=RoyalBlue,urlcolor=Black,linkcolor=RoyalBlue]{hyperref}
\usepackage[nameinlink,capitalize]{cleveref}
\usepackage{titlesec}

\renewcommand{\thesection}{\arabic{section}}
\renewcommand{\thesubsection}{\thesection.\arabic{subsection}}
\renewcommand{\thesubsubsection}{\thesubsection.\arabic{subsubsection}} 

\setcitestyle{super}

\begin{document}

\renewcommand{\figurename}{Fig.} 
\renewcommand{\tablename}{Tab.} 

\title{Cell clusters sense their global shape to drive collective migration}

\author{Joan Térmens}
\affiliation{Departament de Física de la Matèria Condensada, Universitat de Barcelona (UB), 08028 Barcelona, Spain}
\affiliation{Universitat de Barcelona Institute of Complex Systems (UBICS), 08028, Barcelona, Spain}

\author{Irina Pi-Jaumà}
\affiliation{Departament de Física de la Matèria Condensada, Universitat de Barcelona (UB), 08028 Barcelona, Spain}
\affiliation{Universitat de Barcelona Institute of Complex Systems (UBICS), 08028, Barcelona, Spain}

\author{Ido Lavi}
\affiliation{Departament de Física de la Matèria Condensada, Universitat de Barcelona (UB), 08028 Barcelona, Spain}
\affiliation{Center for Computational Biology, Flatiron Institute, 162 5th Ave, New York, NY 10010, USA}

\author{Marija Matejčić}
\affiliation{Institute for Bioengineering of Catalonia (IBEC), 08028, Barcelona, Spain}

\author{Isabela Corina Fortunato}
\affiliation{Institute for Bioengineering of Catalonia (IBEC), 08028, Barcelona, Spain}
\affiliation{Centro de Investigación Biomédica en Red en Bioingeniería, Biomateriales y Nanomedicina (CIBER-BBN), 08028 Barcelona, Spain}

\author{Xavier Trepat}
\affiliation{Institute for Bioengineering of Catalonia (IBEC), 08028, Barcelona, Spain}
\affiliation{Centro de Investigación Biomédica en Red en Bioingeniería, Biomateriales y Nanomedicina (CIBER-BBN), 08028 Barcelona, Spain}
\affiliation{Departament de Biomedicina, Universitat de Barcelona (UB), 08036 Barcelona, Spain}
\affiliation{Institució Catalana de Recerca i Estudis Avançats (ICREA), 08010, Barcelona, Spain}

\author{Jaume Casademunt}
\email{jaume.casademunt@ub.edu}
\affiliation{Departament de Física de la Matèria Condensada, Universitat de Barcelona (UB), 08028 Barcelona, Spain}
\affiliation{Universitat de Barcelona Institute of Complex Systems (UBICS), 08028, Barcelona, Spain}

\begin{abstract}
The collective migration of epithelial groups of cells plays a central role in processes such as embryo development, wound healing, and cancer invasion. While finite cell clusters are known to collectively migrate in response to external gradients, the competing effect of possible endogenous cues is largely unknown.Here, we demonstrate that the polarization of peripheral cells that pull the cluster's edge outward is sufficient to induce and sustain the collective migration of confluent clusters. We use a general continuum model to show that the underlying shape-sensing mechanism is purely mechanical, relying on long-range hydrodynamic interactions and cell-cell alignment forces. As a proof-of-concept, we validate our findings with experiments on monolayer clusters from various cell lines, where we control initial shapes and sizes. The mechanism operates independently of external signals and will generally interfere with them. Specifically, we predict and observe experimentally that it can override collective durotaxis, reversing the direction of migration. Together, our results offer a physical framework for understanding how cell interactions govern the interplay between global shape and collective motion and afford engineering principles for optimal control and manipulation of cell cluster shape and motion.
\end{abstract}

\maketitle 

\section{Introduction} \label{sec:intro}
Collective cell migration in confluent tissues is central to various biological processes, from embryo development to wound healing or the metastasis cascade in cancer progression \cite{Friedl2009, Majumdar2014, Haeger2015}. Yet, while different mechanisms of individual cell motility are largely studied and well-known, the coordination mechanisms that orchestrate collective cell migration in confluent tissues remain poorly understood. Specifically, the case of finite cell clusters with a deformable free boundary undergoing directed migration, studied both \emph{in vivo} \cite{Shellard2021a, Shellard2021b} and \emph{in vitro} \cite{Pallarès2023}, poses additional challenges due to the complex coupling of the free-boundary dynamics and the mechanics of internal cells. In general, collective migration responds to some symmetry-breaking driving field, somewhat mimicking and extending a response that occurs already at the individual cell level, such as in durotaxis \cite{Lo2000, Vincent2013, DuChez2019, Sunyer2016, Sunyer2020, Shellard2021a, Pallarès2023}, chemotaxis \cite{Jin2008, Poukkula2011, Camley2016}, electrotaxis \cite{Cortese2014, Cohen2014, Liu2014, Lyon2019, Prescott2021}, frictiotaxis \cite{Shellard2023}, haptotaxis \cite{Wen2015, King2016, Luo2020, Fortunato2024}, etc. In addition, for finite clusters, global motion may be induced by some internal symmetry breaking, where cells at different locations of the cluster exhibit different properties, such as in local optogenetic stimulation experiments \cite{Rossetti2024}. 

In the absence of external cues or induced symmetry-breaking, however, an epithelial monolayer of identical cells may still be driven by endogenous cues. In the case of a free-boundary cluster, the polarization of cells at the edge that tends to drive local migration outward is a prominent endogenous cue \cite{Poujade2007, Trepat2009, Pérez-González2019}. Consequently, the tissue may spread due to the traction force of the peripheral cells possibly competing with cell-cell contractile forces that tend to favour retraction of the edge. In addition, monolayer edges are known to undergo an active fingering instability that produces complex morphologies \cite{Pérez-González2019,Alert2019a}. Accordingly, spontaneous symmetry breaking of a circular cluster will generate nontrivial shapes in response to random fluctuations. This raises the question of whether such spontaneous (morphological) symmetry breaking can lead to collective migration. 

We note that the outward pulling by peripheral cells does not impose a globally privileged direction. Unlike the driving by external fields, which already operates at the single-cell level, the driving by such endogenous cues is inherently collective as it relies on the shape of the cluster boundary. In terms of a continuum mechanical model, the symmetry is to be broken by the initial condition in an otherwise translationally and rotationally invariant set of equations and boundary conditions. A similar precedent of spontaneous motility due to morphological symmetry breaking was discussed in Refs.\cite{Callan-Jones2008,Blanch-Mercader2013} in the context of cell lamellar fragments\cite{Verkhovsky1998}. However, the physics of cell monolayers, which combines cell-to-cell and cell-to-substrate active forces, together with more complex constitutive equations, is much more complex that of actin gels in thin fragments. 

A previous study of spontaneous migration of tissue clusters reported the formation of the so-called giant keratocytes \cite{Beaune2018}, namely spheroids on a substrate that get globally polarized in a head-tail 3D complex structure reminiscent of individual keratocytes. In a similar spirit to how the study of cell fragments shed new light on the mechanisms of cell motility, our present study of cell monolayer clusters may also be instrumental in understanding the motility of more complex 3D tissue structures. Along this line, we remark that during the initial stages of giant keratocytes, spontaneous symmetry-breaking occurs in a wetting monolayer surrounding the spheroid through a morphological instability of the tissue front, relating directly to the physics of monolayers.

Finally, we remark that the phenomenon of shape-sensing motility discussed here in an idealized setup, is expected to be generic and present in more realistic environments and \emph{in vivo}, coexisting with the effect of other external and internal cues, and thereby alter the patterns of induced migration %in more realistic situations, 
whenever clusters adopt non-trivial shapes or, in general, are allowed to deform. Furthermore, unravelling the interplay between endogenous and exogenous cues opens new possibilities of practical relevance in the design, optimal control, and manipulation of tissues, for instance in the spirit of electrotactic control discussed in Refs.\cite{Cohen2014, Martina-Perez2024}.

\section{Theory of an active moving drop} \label{sec:theory}

Collective cell migration of epithelial cells is often organized at supra-cellular scales \cite{Poujade2007, Friedl2009, Vedula2013, Mayor2016, Ladoux2017, Hakim2017, Alert2020}. Accordingly, a phenomenological continuum description of epithelia, coarse-grained at scales larger than the cell size \cite{Jülicher2018, Brückner2024}, has been instrumental to provide insights into a variety of biological processes \cite{Arciero2011, Lee2011a, Lee2011b, Marel2014, Recho2016}, and has shown remarkable predictive power in a variety of applications \cite{Blanch-Mercader2017a, Blanch-Mercader2017b, Pérez-González2019, Alert2019a, Alert2019b, Pallarès2023}.

\subsection{Center-of-mass velocity} \label{sec:center_of_mass}

We will describe the dynamics of a finite monolayer cluster as the time-evolution of its 2D domain $\Omega(t)$, like that in \cref{fig:evolutions}a,b, where a velocity field $\mathbf{v}$ is defined. This entails a free-boundary problem in which the evolution of the boundary $\partial\Omega(t)$ is defined by the continuity condition $v_n=\mathbf{v}\cdot\mathbf{\hat{n}}$, where $\mathbf{\hat{n}}$ is the unit (outward) normal vector along the boundary. We are primarily interested in the geometric center of mass $\mathbf{R_{CM}}$ of this planar shape, i.e. the centroid, and more specifically on its velocity $\mathbf{V_{CM}} = \mathbf{\dot{R}_{CM}}$. We remark that this aerial center of mass of the planar shape depends solely on the boundary, and thus differs from the physical center of mass, which would take into account the variations of the nonuniform thickness of the monolayer $h(\mathbf{r},t)$. In the Supplementary Material, we discuss how the variations of $\mathbf{\hat{n}}$ are related to the compressibility of the 2D velocity $\mathbf{v}$, and we show that the velocity of the geometric center of mass is given by

\begin{equation} 
\mathbf{V_{CM}}= \frac{1}{A} \left[\int_\Omega \mathbf{v}\;dS + \int_\Omega  \mathbf{r}\; \left(\mathbf{\nabla}\cdot\mathbf{v} - \left<\mathbf{\nabla}\cdot\mathbf{v} \right>\right)\;dS \right], 
\label{eq:kinematics}
\end{equation}

where $\left< ... \right> \equiv \frac{1}{A}\int_{\Omega} ...dS$ and $A\equiv\int_{\Omega}dS$ is the total area, so $\mathbf{R_{CM}} =\left< \mathbf{r} \right>$, and $\dot{A}=\left<\mathbf{\nabla}\cdot\mathbf{v} \right>$. The second term in the right-hand side (rhs) of \cref{eq:kinematics} accounts for the inhomogeneous distribution of the local spreading rate $\mathbf{\nabla}\cdot\mathbf{v}$, and vanishes for an incompressible 2D flow (i.e. $\mathbf{\nabla}\cdot\mathbf{v}=0$), but also for one with a uniform spreading rate (i.e. $\mathbf{\nabla}\cdot\mathbf{v}=\alpha(t)$).  

\cref{eq:kinematics} is purely kinematic, and needs to be complemented with an explicit dynamical model to determine $\mathbf{v}$. With great generality, we assume that such model will also involve another vector field $\mathbf{p}$ to account for cell polarization, so that cells can exert traction forces on the environment, and an equation of state $\Pi(\rho)$ that relates pressure $\Pi$ to density $\rho$. Then for an arbitrary stress tensor $\mathbf{\sigma}$, and in the absence of inertia, we assume the generic force balance equation

\begin{equation}
\mathbf{\nabla}\cdot\mathbf{\sigma} - \mathbf{\nabla}\Pi = \xi \; \mathbf{v} - \zeta_{\text{i}} \;\mathbf{p}, 
\label{eq:external}
\end{equation}

where the rhs accounts for the external force density exerted through the contact with the environment, and relates to the experimentally measured force per unit area when multiplied by $h$, as discussed in Ref.\cite{Pérez-González2019}. Based on symmetry arguments and very general considerations \cite{Oriola2017}, the contact force with the substrate is split into a passive friction proportional to velocity, with a friction coefficient $\xi$, and an active traction force, referred hereinafter simply as `traction force', proportional to polarization $\mathbf{p}$ and whose maximal magnitude is taken to be $1$. Thus, the traction coefficient $\zeta_{\text{i}}$ yields the maximal active traction force cells can exert when they are fully polarized. Imposing stress-free boundary conditions, and allowing for a Young-Laplace normal force due to surface tension $\gamma$,

\begin{equation}
    (\mathbf{\sigma}-\Pi {\bold{1}})\colon \mathbf{\hat{n}} \mathbf{\hat{n}} = -\gamma (\nabla\cdot \mathbf{\hat{n}}),
\end{equation}

where $\bold{1}$ is the identity tensor, integration of \cref{eq:external} yields

\begin{equation}
    \int_\Omega \mathbf{v} \;dS = \frac{\zeta_{\text{i}}}{\xi} \int_\Omega \mathbf{p} \;dS.
    \label{eq:v-p}
\end{equation}

Then, combining \cref{eq:kinematics} and \cref{eq:v-p} we rewrite \cref{eq:kinematics} as 
  
\begin{equation}
    \mathbf{V_{CM}} = \mathbf{I}_\Omega^T + \mathbf{I}_\Omega^S,
\end{equation}
 
with

\begin{align}
    \mathbf{I}_\Omega^T &\equiv  \frac{1}{A} \frac{\zeta_{\text{i}}}{\xi} \int_\Omega \mathbf{p}\;dS, 
    \label{eq:IT} \\
    \mathbf{I}_\Omega^S  &\equiv \frac{1}{A}\int_\Omega \left( \mathbf{r}-\mathbf{R_{CM}}\right) \mathbf{\nabla}\cdot\mathbf{v}\;dS. \label{eq:IS}
\end{align}

We refer to $\mathbf{I}_\Omega^T$ as the traction integral and to $\mathbf{I}_\Omega^S$ as the spreading integral. $\mathbf{I}_\Omega^T$ embodies the propulsion caused by a net active traction force, which pulls the center of mass as if it were an overdamped particle with an effective friction coefficient $A\xi$, that is, as if the domain $\Omega$ were moving like a rigid solid. $\mathbf{I}_\Omega^S$, re-expresses the second term in the rhs of \cref{eq:kinematics} as the dipolar moment of the local expansion rate.

We now address the dynamics of $\mathbf{p}$. Following previous studies for spreading monolayers\cite{Blanch-Mercader2017b, Alert2019a,Pérez-González2019}, we model the tissue as a 2D active (polar) nematic fluid in the isotropic phase, with cell-cell polarity alignment interactions with an effective free energy

\begin{equation}
    F = \int \left[ \frac{1}{2} a \;\mathbf{p}\cdot\mathbf{p} + K\; \nabla\mathbf{p}\colon\nabla\mathbf{p} \right]  \;dS .
    \label{eq:Frank_energy}
\end{equation} 

The first term expresses the energy cost of polarization so that $\mathbf{p}=0$ yields the equilibrium energy. The second term penalizes the misalignment of polarized regions. 

The characteristic endogenous cue for epithelial monolayers is enforced by the boundary condition $\mathbf{p}|_{\partial \Omega} = \mathbf{\hat{n}}$, imposing that edge cells are  fully polarized, pointing outwards perpendicularly. This implies that the free-energy minimum will correspond to a polarization field that decays from $\mathbf{\hat{n}}$ at the boundary to zero in the interior with a characteristic decay length $L_c \equiv \sqrt{K / a}$ that balances both terms. $L_c$ is often referred to as the nematic correlation length, or the nematic screening length, and the peripheral region as the polarized boundary layer. $L_c$ thus defines the penetration length of the endogenous cue associated with the edge cells, which typically amounts to a few cell lengths.

The dynamics of $\mathbf{p}$ is usually much faster than that of $\mathbf{v}$, so a common assumption is that the polarization dynamics is purely relaxational and decoupled from $\mathbf{v}$ \cite{Pérez-González2019}, that is,

\begin{equation}
    \frac{\partial \mathbf{p}}{\partial t} \approx - \frac{1}{\gamma}\frac{\delta F}{\delta \mathbf{p}} \approx 0,
    \label{eq:quasi_static}
\end{equation}

which typically holds when the fluid viscosity $\eta$ is much larger than the rotational viscosity $\gamma$. 
This implies that $\mathbf{p}$ adopts instantaneously the equilibrium configuration corresponding to the current boundary, as the solution of the screened Laplace equation

\begin{equation}
    \left( L_c^2 \nabla^2 - 1\right) \mathbf{p}= 0.
    \label{eq:quasi_static_solution}
\end{equation}

Under these conditions, the polarization field and hence the traction integral $\mathbf{I}_\Omega^T$, is determined solely by the shape of the domain $\Omega$ and the scale $L_c$. 

The nematic screening length $L_c$ sets a range of nonlocality for the polarization field, that is, the range beyond which this field becomes decorrelated. We note that $L_c$ must be finite compared with a characteristic system size $L$ to allow for a finite value of $ \mathbf{I}_\Omega^T$ since  $\lim_{L_c/L\rightarrow 0} \mathbf{I}^T_\Omega =0$ even if the limit is taken keeping $L_c \zeta_{\text{i}} =\text{const}$. This is so because $\int_{\partial \Omega} \mathbf{\hat{n}} ds = 0$, where $s$ is the arclength, regardless of the domain's shape. If $L_c/L$ is finite, then a finite value of the traction integral may be expected provided the curvature changes on the scale $L_c$, and the global shape defines a head-tail polarity. 

Assuming a viscous-like constitutive equation for the tissue with a viscosity $\eta$, we expect yet another important source of nonlocality set by the hydrodynamic interactions, with a characteristic screening length $\lambda \equiv \sqrt{\eta/\xi}$. While $L_c$ is typically smaller than $L$, $\lambda$ will typically be larger or comparable to $L$ ensuring mechanical force transmission throughout the cluster, which ultimately confers shape sensitivity to the cluster as a whole. 

\subsection{Sustained motion in a minimal model}\label{sec:model_equations}

To see whether a finite $\mathbf{V_{CM}}$ for a given shape may be sustained and/or amplified by a positive feedback, we need to specify a dynamical model. For simplicity, we will focus on a minimal model that has been shown to capture the basic physics of spreading epithelia in many experiments, and for which good parameter estimates are available (see Refs. \cite{Blanch-Mercader2017b, Pérez-González2019, Alert2019a, Alert2019b} for details).
We take a fluid constitutive equation of the form

\begin{equation}
    \mathbf{\sigma} = \eta \; (\nabla \mathbf{v} + \nabla \mathbf{v}^T) - \zeta \;\mathbf{p} \mathbf{p},
    \label{eq:constitutive}
\end{equation}

which combines viscous stress and active contractile stress. The latter is quantified by the contractility parameter $\zeta<0$. We assume that the 2D fluid is highly compressible (it can accommodate local 2D compression/expansion by changing $h$ without a significant change of pressure), so $\nabla \Pi \approx 0$, and hence the pressure drops from the formulation. Surface tension is also neglected for simplicity. We note that shape-sensing motility arises independently of both simplifications. Then, combining \cref{eq:constitutive} with \cref{eq:external}, we obtain 

\begin{equation}
    \lambda^2 \left(\nabla^2 \mathbf{v}  
    + \nabla (\mathbf{\nabla}\cdot \mathbf{v})\right)
    - \mathbf{v} = \mathbf{P}
    \label{eq:velocity}
\end{equation}

with

\begin{equation}
    \mathbf{P} \equiv   -\frac{\zeta_{\text{i}}}{\xi}  \left(\mathbf{p} +\ell_a \mathbf{\nabla}\cdot (\mathbf{p}\mathbf{p}) \right)
\end{equation}

where we define the active length scale $\ell_a \equiv |\zeta| / \zeta_{\text{i}} $,  which measures the relative importance of the two active forces, namely the traction forces $\zeta_{\text{i}} \mathbf{p}$ on the substrate and the contractile cell-cell forces $\zeta \mathbf{\nabla}\cdot \mathbf{p}\mathbf{p}$. \cref{eq:quasi_static_solution} and \cref{eq:velocity} together with the stress-free boundary condition $\mathbf{\sigma}\colon \mathbf{\hat{n}} \mathbf{\hat{n}} =0 $ and the kinematic condition $v_n=\mathbf{v}|_{\partial \Omega} \cdot \mathbf{\hat{n}}$, define the evolution of the domain $\Omega(t)$. Taking the divergence and the curl on \cref{eq:velocity}, we split it into two equations for the local expansion rate $\chi\equiv \mathbf{\nabla}\cdot \mathbf{v}$ and the vorticity $\omega \equiv (\mathbf{\nabla} \times \mathbf{v}) \cdot \mathbf{\hat{k}}$, which both take the form of a screened Poisson equation

\begin{align}
    \left( 2\lambda^2 \nabla^2 - 1\right) \chi &= 
    \mathbf{\nabla}\cdot \mathbf{P}, \label{eq:div}  \\
    \left( \lambda^2 \nabla^2 - 1\right) \omega &= (\mathbf{\nabla} \times \mathbf{P}) \cdot \mathbf{\hat{k}}. \label{eq:curl}
\end{align}

\begin{figure}[htb!]
    \centering
    \includegraphics[width=\columnwidth]{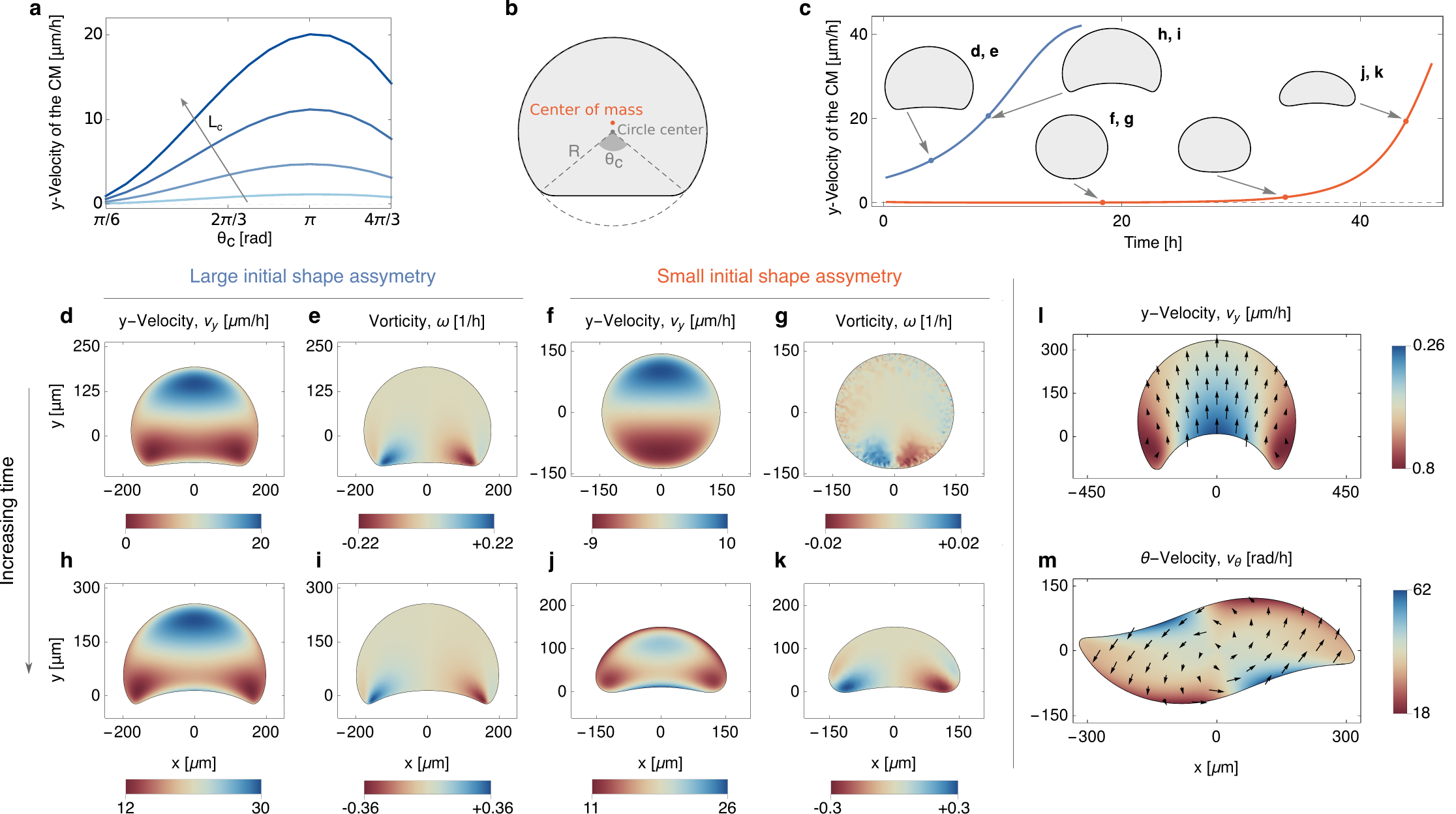}
    \caption{\textbf{Temporal evolutions for cell clusters with different shapes and parameters.} \textbf{a,} y-component of the CM velocity for cutted-circles (initial cluster shapes of the simulations) depending on its cut and for varying nematic lengths. The maximum of the velocity is reached for big nematic lengths and around the semicircle ($\theta_c = \pi \,\text{rad}$). \textbf{b,} Sketch of a cutted-circle with rounded corners showing its radius and $\theta_c$. \textbf{c,} Temporal evolutions of two clusters with different initial shapes. The plot shows the evolution of the y-component of the center of mass velocity together with some contours of the clusters at different stages of their evolutions. The initial sizes of the clusters were selected so that the simulations start close to the spreading transition (see \cref{tab:params_shape}), and the $L_c$ was set to $L/4$. The other parameters are the following: $\eta = 50 \,\text{MPa·s}$, $\xi = 0.1 \,\text{kPa·s}/\mu\text{m}^2$, $\zeta_{\text{i}} = 0.1 \,\text{kPa}/\mu\text{m}$ and $\zeta = -20 \,\text{kPa}$. \textbf{d-k,} Plots of the y-component of the velocity and vorticity for the contours at subfigure \textbf{c}. Subfigures \textbf{d} and \textbf{e} corresponds to an early state of a cluster with a large front-rear asymmetry, while \textbf{f} and \textbf{g} show that even very small asymmetries are sufficient to trigger the cluster propulsion. Subfigures \textbf{h} to \textbf{k} show advanced stages of the clusters in the above row. \textbf{l,} Y-component of the velocity for an incompressible cluster. The arrows show the direction of the velocity and are scaled according to its magnitude. \textbf{m,} Angular velocity for a chiral cluster. The parameters of the shapes are specified at \cref{tab:params_shape} and the remaining ones are the same as in \textbf{c}.}
    \label{fig:evolutions}
\end{figure}

We see that the polarized boundary layer contains the source $\mathbf{\nabla}\cdot \mathbf{P}$ for the spreading rate, and $(\mathbf{\nabla} \times \mathbf{P}) \cdot \mathbf{\hat{k}} $ for the vorticity. \cref{eq:div} controls the spreading transition reported in Ref.\cite{Pérez-González2019}, according to which there is a crossover at a given system size between global spreading ($\dot{A} >0$) when traction dominates and global retraction ($\dot{A} < 0$) when contractility dominates. On the other hand, \cref{eq:curl} tells us that, for a non-circular shape and with a finite $L_c$, $\mathbf{p}$ is not radial within the boundary layer, thus containing a source of vorticity. Thus, the effect of having a net traction force can be traced to the appearance of vorticity in the flow. Within the polar axisymmetry, vorticity will have opposite signs on both sides of the domain $\Omega$, which will ultimately drive the global motion of the cluster. For a chiral shape, instead, there would be a dominant sign of vorticity, leading to the rotation of the cluster (see \cref{fig:evolutions}l). 

In \cref{fig:evolutions}c-k, we show examples of sustained evolutions of clusters with different initial shapes. The parameters are taken within realistic ranges, with
the cluster size, $L$, taken close to the spreading transition so that the area does not change significantly with time. In \cref{fig:evolutions}f,g,j,k the initial shape is almost perfectly circular, with just a minor flattening at the bottom, demonstrating the generic (nonlinear) instability of circular configurations to traveling modes.

\subsection{Collective motility modes} \label{sec:modes}

\begin{figure}[htb!]
    \centering
    \includegraphics[width=\columnwidth]{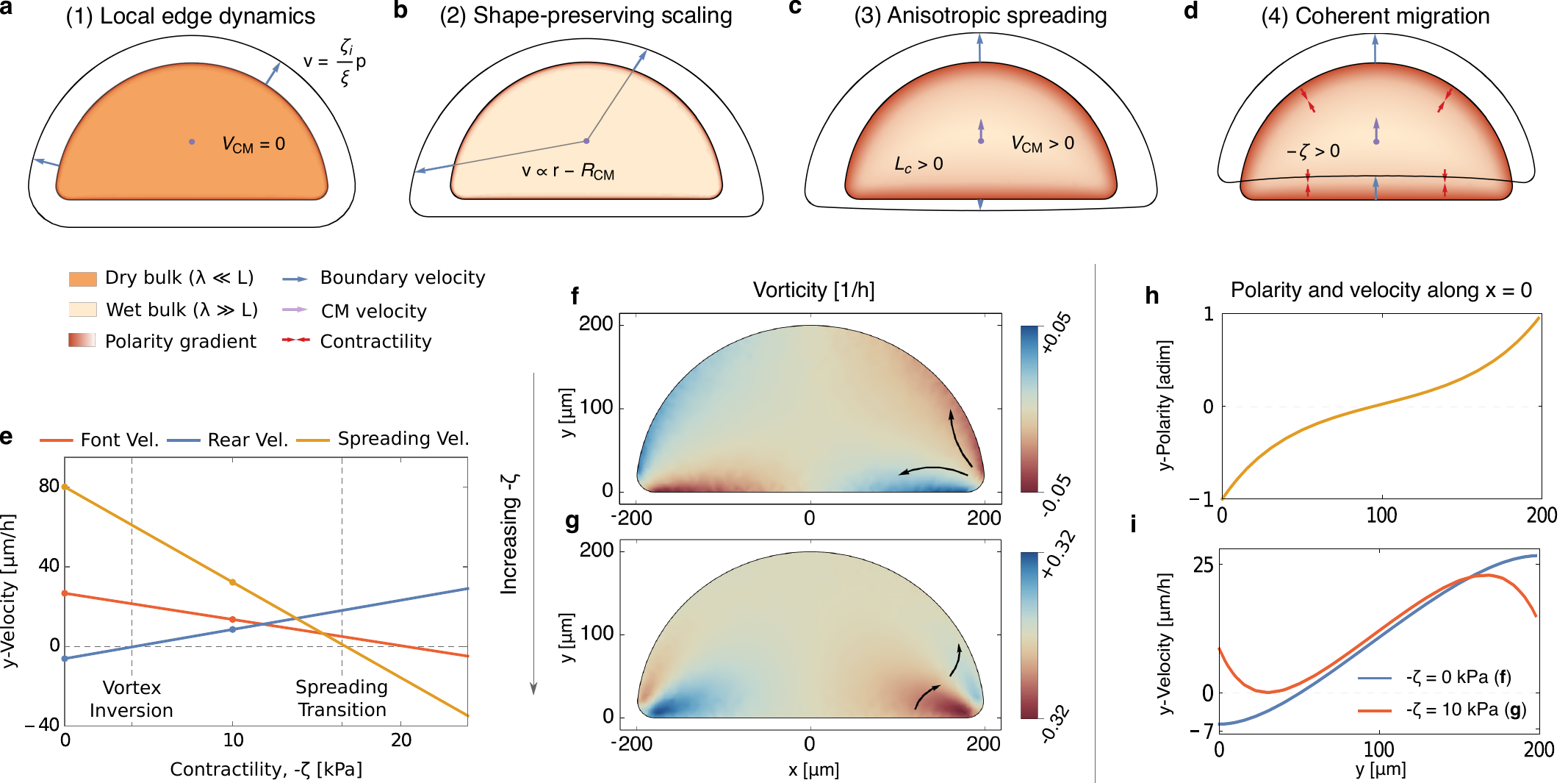}
    \caption{\textbf{Schematic representation of the motility modes and characterization of the transition between anisotropic spreading and coherent migration.} \textbf{a-d,} Schematic illustration of the four motility Modes for a prototypic semicircular shape. Modes 2 to 4 present non-local dynamics, Modes 3 and 4 generate a net velocity of the center of mass and Mode 4 shows the retraction of the rear. \textbf{e,} The plot shows how the front, rear, and spreading velocities change for increasing contractility, $-\zeta$. The dashed lines indicate the level of zero-velocity as well as the vorticity inversion (-$\zeta \simeq \zeta_{\text{i}} L_c$) and the spreading transition ($\sqrt{A}\langle \nabla \cdot \mathbf{v}\rangle = \dot{A}/A = 0$). \textbf{f, g,} show the vorticity of the clusters at two different contractilites: $-\zeta = 0$ kPa (\textbf{f}, top), and $-\zeta = 10$ kPa (\textbf{g}, bottom). The inversion of the vorticity upon increasing the contractility indicates a transition between Mode 3 (anisotropic spreading, \textbf{c}) and Mode 4 (coherent migration, \textbf{d}). \textbf{h,} Plot of the y-component of the polarity along the symmetry axis ($x=0$) for the left clusters, \textbf{f} and \textbf{g}. \textbf{i,} Plot of the y-component of the velocity along the symmetry axis ($x=0$) for the simulations at \textbf{f} and \textbf{g}. The plot shows how, at low contractilities (case \textbf{f}, in blue), the velocity profile follows the polarity and changes sign at the rear, while at higher contractilities, the rear velocity turns positive, and thus we observe a rear retraction and coherent migration.}
    \label{fig:modes}
\end{figure}

The minimal model at hand can be made dimensionless by rescaling variables by the characteristic time scale $\eta/(\zeta_{\text{i}} L_c)$, length scale $L$, and stress scale $\zeta_{\text{i}} L_c$. The parameter space is then best organized by considering the relative size of the four relevant length scales $L$, $L_c$, $\lambda$, and $\ell_a$. We will restrict ourselves to realistic situations where $L_c < L < \lambda$. Here we define four phenomenological scenarios one may conceive a priori and and discuss how they relate to the model parameters. A schematic illustration for a prototypical shape can be seen in \cref{fig:modes}a-d.

\begin{itemize}
    \item Mode 1 ($\mathbf{V_{CM}}=0$). \emph{Local edge dynamics}: the normal velocity is constant along the boundary. The shape is not preserved.

    \item Mode 2 ($\mathbf{V_{CM}}=0$). \emph{Shape-preserving scaling}: $\mathbf{\nabla}\cdot\mathbf{v}=\text{const}$. The shape is preserved as the time evolution corresponds to a global rescaling.

    \item Mode 3 ($\mathbf{V_{CM}} \neq 0$). \emph{Anisotropic spreading}: $\mathbf{\nabla}\cdot\mathbf{v} \neq \text{const}.$ In the polarized boundary layer, $\mathbf{v}$ and $\mathbf{p}$ are aligned, $\mathbf{v}\cdot\mathbf{p} > 0$ (see \cref{fig:modes}h,i in blue). There is no global alignment of the velocity field ($\mathbf{v}\cdot\mathbf{V_{CM}}>0$ in the front and $\mathbf{v}\cdot\mathbf{V_{CM}}<0$ in the rear). The shape evolution tends to slow down $\mathbf{V_{CM}}<0$. 
    
    \item Mode 4 ($\mathbf{V_{CM}} \neq 0$). \emph{Coherent migration}: the cluster moves as a whole with retraction of the rear, where velocity is reversed, $\mathbf{v}\cdot\mathbf{p} < 0$ (see \cref{fig:modes}h,i in red). There is a global alignment of the velocity field, with $\mathbf{v}\cdot\mathbf{V_{CM}}>0$ almost everywhere. The shape evolution yields a positive feedback, speeding up $\mathbf{V_{CM}}<0$.
\end{itemize}

Mode 1 is obtained in the local limit where both $L_c \ll L$ and $\lambda \ll L$ (`dry' limit). Then, all cells along the edge behave equally, insensitive to shape or size of the tissue cluster, with a constant normal velocity $v_n=\zeta_{\text{i}}/\xi$. Cells in the interior have $\mathbf{v}\simeq 0$. The total traction force vanishes and the center of mass does not move, even though the shape and size change.

Mode 2 is obtained in the `wet' limit, with $\lambda \gg L$ keeping $L_c \ll L$. The problem is fully nonlocal regarding the hydrodynamic interactions, but it is local for the polarity. In this case, the velocity approaches $\mathbf{v} = \frac{1}{2}\alpha (\mathbf{r}-\mathbf{R_{CM}})$, which solves $\mathbf{\nabla}\cdot\mathbf{v}=\alpha$. Cells are sensitive to the global shape, as their behavior depends on their relative position to the center of mass. The shape is preserved under the evolution, but the center of mass does not move. 

Mode 3 is obtained in the `wet' limit but now with finite $L_c$ (i.e. $L_c \lesssim L$). This is the simplest case where we have a finite velocity of the center of mass, and will be studied experimentally in the sections below. The cells now respond to long-range hydrodynamic and alignment interactions. The velocity profile along the edge is non-trivial and sensitive to the global shape. 

Mode 4 is obtained when tissue contractility is large enough to be close to the spreading transition \cite{Pérez-González2019}, so the change of area is not significant.  In this case, the increase of contractility causes the vorticity distribution to reverse at the corners with respect to Mode 3, thus inducing the retraction of the rear front. \cref{fig:modes}e-i illustrates the velocity reversal phenomenon as contractility increases, which defines the transition between Mode 3 and Mode 4, as the rear velocity follows the inversion of the vorticity distribution. We remark that the coherent motion of the cluster in Mode 4 is not a flocking phenomenon as described in the context of polarized tissues \cite{Giavazzi2018}, since the velocity field is globally oriented but the polarization field is not. 

Mode 4 can also be achieved in the case of an incompressible flow ($\mathbf{\nabla}\cdot\mathbf{v}=0$), showing that it relies essentially on the fact that the area is not changing significantly, rather than due to an effect of contractility, as shown in \cref{fig:evolutions}m. This also shows that this coherent migration mode is quite general, spanning the whole range of compressibility from zero to infinity. 

\subsection{Friction-driven propulsion mechanism}

\begin{figure}[htb!]
    \centering
    \includegraphics[width=0.9\columnwidth]{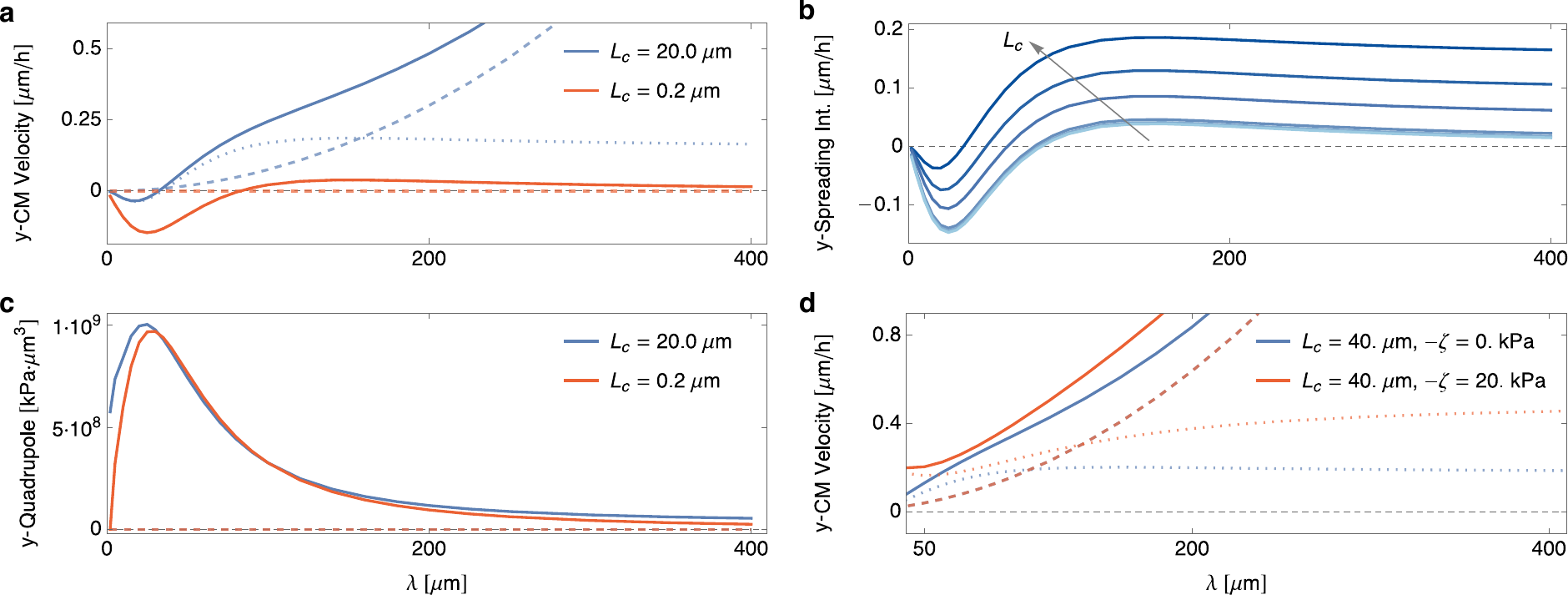}
    \caption{\textbf{Comparison between traction and friction-driven propulsion mechanisms.} \textbf{a,} Decomposition of the y-component of the center of mass velocity (solid lines) into both traction (dashed) and spreading (dotted) integrals as a function of the screening length ($\lambda$). The blue lines show the results of simulations with $L_c=20\,\mu\text{m}$ which are dominated by the traction integral, while the red ones show $L_c=0.2\,\mu\text{m}$, which closely follow the spreading one. The limit of the red curve for low $\lambda$ approximates Mode 1 (\cref{fig:modes}a), while its limit for large $\lambda$ approximates  Mode 2 (\cref{fig:modes}b). The regions with large $\lambda$ of the blue curve correspond to Mode 3 (\cref{fig:modes}c). \textbf{b,} Variation of the spreading integral for different nematic lengths along $\lambda$. The darker curve corresponds to $L_c = 20 \,\mu\text{m}$, while the lighter one corresponds to $L_c = 0.2 \,\mu\text{m}$. \textbf{c,} The plot shows the $yyy$ component of the force quadrupole, $Q_{yyy} = \int_\Omega (y-Y_{CM})^2f_y \,\mathrm{d}S$ (were $f_y$ stands for the y-component of the force $\mathbf{f}$), for both the traction force (dashed lines) and the friction force (solid lines) for two nematic lengths: $L_c=20\,\mu\text{m}$ (blue) and $L_c=0.2\,\mu\text{m}$ (red). Note that the traction quadrupoles are null and that the maximum of the friction quadrupole corresponds to the maximum magnitude of the spreading integral. \textbf{d,} Plot analogous to \textbf{a} that shows the effect of the contractility. The blue lines show the results of simulations with $L_c=40\,\mu\text{m}$ and no contractility, while the red ones show $L_c=40\,\mu\text{m}$ with $-\zeta = 20\,\text{kPa}$, which depicts Mode 4 (\cref{fig:modes}d) at the regions with large $\lambda$. Note that the spreading integral (dotted) increases upon increasing the contractility and that the traction integral (dashed) also corresponds to the CM velocity of an incompressible cluster, which would be slower than its compressible counterpart due to its null spreading integral. The maximal active stress is kept constant to $\zeta_{\text{i}} L_c = 1 \,\text{kPa}$ along the different simulations and the contractility is null. Cluster size and shape is specified at \cref{tab:params_shape}.}
    \label{fig:friction}
\end{figure}

For relatively small friction forces, $\lambda /L \gg 1$, the traction integral $\mathbf{I}_\Omega^T$ will typically dominate, and the spreading integral $\mathbf{I}_\Omega^S$ introduces only small corrections. However, the situation becomes more complex when $\lambda /L \lesssim 1$ but $L_c/L \ll 1 $ (with finite $L_c\zeta_{\text{i}}$), implying that $\mathbf{I}_\Omega^T$ is very small. In \cref{fig:friction}a,b we show the decomposition of $\mathbf{V_{CM}}$ in the two contributions for the semicircular shape discussed in \cref{fig:modes}, as a function of $\lambda$. While $\mathbf{I}_\Omega^T$ in the $y$-axis is always positive and monotonically increasing with $\lambda$, $\mathbf{I}_\Omega^S$ is non-monotonic and changes sign in the same axis. For small enough $\lambda$ and $L_c$ it overcomes the traction integral contribution producing a $\mathbf{V_{CM}}$ in the opposite direction. In this figure we also visualize the predictions for Mode 1 and Mode 2 when $\lambda \rightarrow 0$ and  $\lambda \rightarrow \infty$, for $L_c/L \ll 1$. In \cref{fig:friction}b we also see that $\mathbf{I}_\Omega^S$ remains finite in the limit $L_c\rightarrow 0$ (keeping $L_c\zeta_{\text{i}}$ finite), showing that a finite $\mathbf{V_{CM}}$ remains in the limit when the total traction force (i.e. the traction monopole) vanishes ($\mathbf{I}_\Omega^T \rightarrow 0$). It can be shown analytically that the traction quadrupole \cite{Rossetti2024} vanishes in this limit too, irrespective of the shape. Therefore, our analysis identifies a propulsive mechanism due to friction, essentially associated with the friction force quadrupole, since the friction monopole will  vanish together with the traction one due to force balance (see \cref{fig:friction}c). Finally, \cref{fig:friction}d shows the effect of contractility on the spreading integral.

\subsection{Spontaneous motility vs durotaxis}
\label{sec:spont_motility}

As an example of how the shape-sensing motility interacts with external cues, we consider how the shape of a 2D cluster influences its speed along a stiffness gradient.
The effect of the inhomogeneous environment is naturally incorporated in the model, by introducing spatially varying traction and friction coefficients, $\zeta_{\text{i}}(\mathbf{r})$ and $\xi(\mathbf{r})$. The effect on the traction integral, which is dominant in the typical parameter regimes relevant to experiments on tissue durotaxis \cite{Pallarès2023}, is given by

\begin{equation}
    \mathbf{I_{\Omega}^T}= \frac{1}{A}   \int_\Omega \frac{\zeta_{\text{i}}(\mathbf{r})}{\xi(\mathbf{r})}\;\mathbf{p(\mathbf{r})}\; dS,
\end{equation}

for the same polarization field $\mathbf{p(\mathbf{r})}$ as in the homogeneous case. For instance, for a constant traction gradient and constant friction, with $\zeta_{\text{i}}(y)= \zeta_{\text{i}}^0 + \zeta_{\text{i}}'(y-y_{\text{CM}})$, we have 

\begin{equation}
    \mathbf{I_{\Omega}^T}= \frac{1}{A\xi} \left[ \zeta_\text{i}^0  \int_\Omega \mathbf{p(\mathbf{r})}\; \mathrm{d}S + \zeta_\text{i}'\int_\Omega (y-y_\text{{CM}}) \mathbf{p(\mathbf{r})}\; \mathrm{d}S   \right].\label{eq:IT_durotaxis}
\end{equation}

In the second integral, the external traction gradient gives rise to the dipolar moment of $\mathbf{p}$, which is nonzero even for a circular shape---unlike the first integral, which is the monopole. For an asymmetric polar shape, both integrals may contribute, possibly in opposite directions.

\begin{figure}[htb!]
  \centering
  \includegraphics[width=0.5\columnwidth]{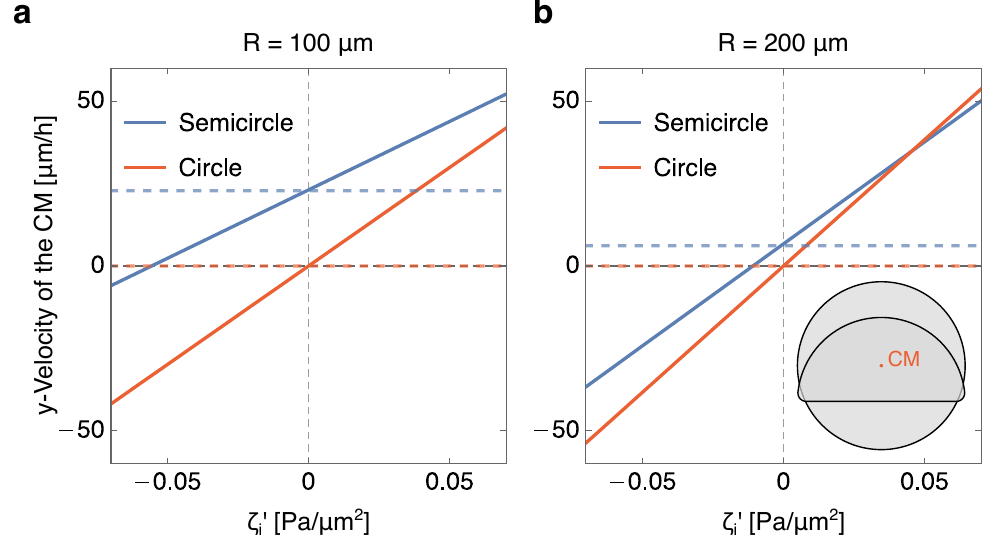} 
  \caption{\textbf{Comparison between durotaxis and shape sensing effects.} Both plots show the value of the y-component of the velocity of the center of mass taking into account durotactic effects, as a function of the durotactic gradient $\zeta_{\text{i}}'$. The solid blue lines correspond to the velocity of a semicircle and the solid red ones, to a circle; while the dashed lines indicate the value of the monopolar term of the traction integral, \cref{eq:IT_durotaxis}, in each of the four cases. The clusters of the right plot have a radius of $R=100\,\mu\text{m}$ and the ones on the left, $R=200\,\mu\text{m}$. The other parameters are $L_c = 25 \,\mu\text{m}$, $\eta = 20 \,\text{MPa·s}$, $\xi = 0.1 \,\text{kPa·s}/\mu\text{m}^2$, $\zeta_{\text{i}}^{0} = 0.1 \,\text{kPa}/\mu\text{m}$ and null contractility, and were adapted from \cite{Pallarès2023,Alert2019b}. In all four different cases the traction integral dominates over the spreading integral, so $\mathbf{V_{CM}} \approx \mathbf{I_{\Omega}^{T}}$.}
  \label{fig:durotaxis}
\end{figure}

For the full computation of $\mathbf{V_{CM}}$ we need to numerically solve $\mathbf{v}$. In \cref{fig:durotaxis} we compare $\mathbf{V_{CM}}$ of clusters with circular and semicircular shapes in the case of a constant friction coefficient and a constant traction-gradient $\zeta_{\text{i}}'$. Within realistic values of stiffness gradients in experiments, we see that the shape-sensing effect on motility is comparable to that of durotaxis. It yields a significant enhancement of the durotactic velocity when the polar asymmetry has the appropriate orientation. Conversely, it may drive motion in the opposite direction, which may overcome durotaxis. The difference is less pronounced for larger clusters, for which the durotactic effect is stronger, and the shape-induced motility is weaker.

\section{Experiments}
\label{sec:experiments}

\begin{figure}[htb!]
  \centering
  \includegraphics[width=\columnwidth]{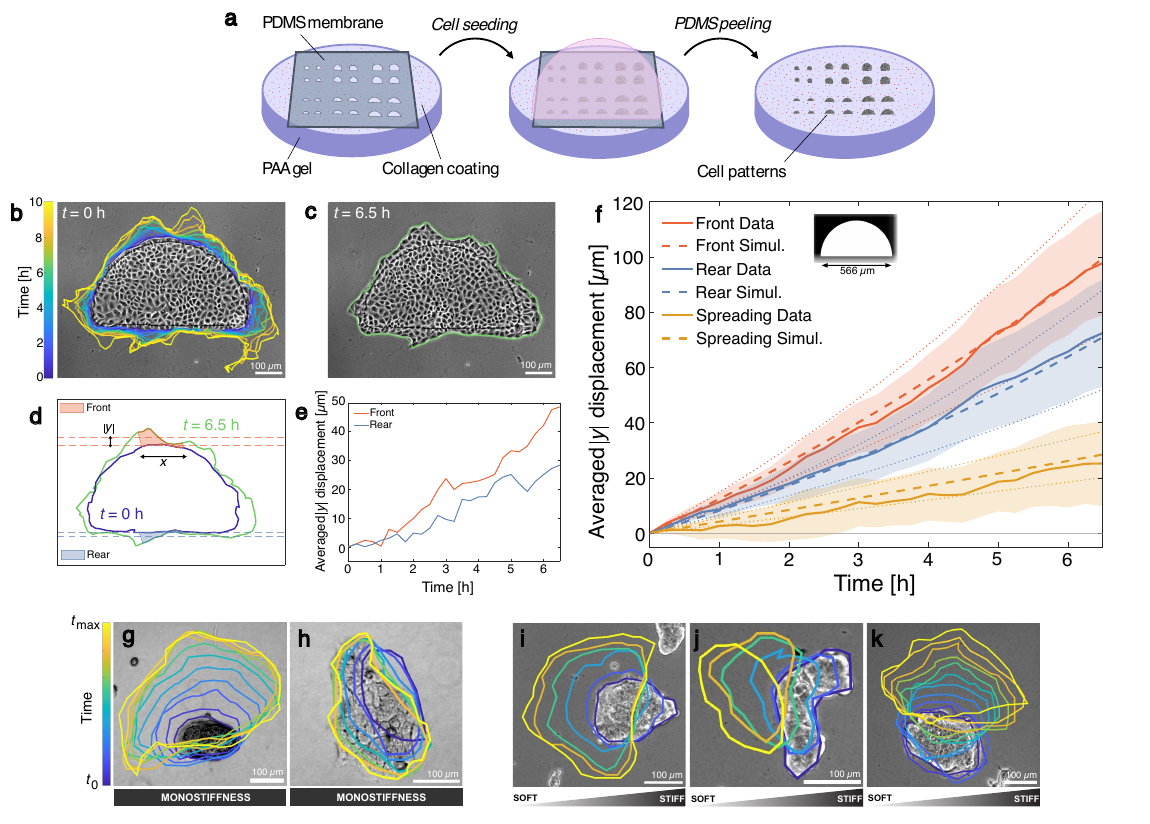} 
  \caption{\textbf{Systematic results for the anisotropic spreading mode and initial evidence of the coherent migration mode.} \textbf{a,} Experimental PDMS membrane patterning assay to obtain patterned monolayers of the desired shapes, adapted from Refs.\cite{Serra-Picamal2012, Pérez-González2019}. Red dots represent the collagen coating and PAA gels have a stiffness of 5 kPa. Each step from the protocol can be found in the Methods section. Large semicircular shapes (C3 in \cref{fig:experiments_patterns}) correspond to panels \textbf{b-f}. \textbf{b,c,} Phase-contrast images of a representative MDCK monolayer, showing the first time point superposed with the evolution of the segmented mask boundaries, where lines are shown every hour (\textbf{b}), and the $t=6.5$ h time point (\textbf{c}). \textbf{d,e,} Boundaries of the monolayer at $t=0$ h and $t=6.5$ h and shaded area (\textbf{d}), used for computing the averaged $\abs{y}$ displacement of the front and the rear over time (\textbf{e}). \textbf{f,} Theoretical results (dashed lines) of the temporal evolution of the front (red), rear (blue), and spreading (yellow, front$-$rear) averaged displacements, which best fit the experimental data for the big semicircles of MDCK cells. Continuous lines are the experimental mean ($N=7$), and the shaded area indicates the Standard Error of the Mean (SEM). The model parameters, $L_c = 35$ $\mu$m and $\eta = 6.25$ MPa$\cdot$s, are selected from the theoretical curves that best fit the experimental data. Dotted lines indicate the lower and upper bounds of the theoretical fits, plotted with $L_c = 25$ and $45$ $\mu$m, respectively. Individual samples yielding to this plot are shown in \cref{fig:experiments_supp_MDCKC3}. \textbf{g-k,} Some examples of unpatterned A431 clusters showing coherent migration, with the evolution of the manually segmented mask boundaries, where lines are shown every hour ($t_{\text{max}} = 10,6,5,5$ and 10 h, from left to right). In \textbf{g} and \textbf{h}, PAA gels have a uniform stiffness of 30 kPa. In \textbf{i-k,} PAA gels have a stiffness gradient increasing towards the right, but since the clusters move towards the left (\textbf{i} and \textbf{j}) or upwards (\textbf{k}), the stiffness is not directing the motion.}
  \label{fig:experiments}
\end{figure}

As a proof-of-concept, we designed a series of experiments to test the motility of monolayer clusters in two lines of cells. We systematically explored a variety of prototypical shapes and sizes for which our numerical simulations predicted a significant center-of-mass velocity. To confine  monolayers to desired initial states, we used polydimethylsiloxane (PDMS) membranes with designed empty regions, ensuring front-rear asymmetry (\cref{fig:experiments_patterns}, Methods). These membranes were placed on top of polyacrylamide (PAA) gels to exploit their linear elastic properties \cite{Storm2005}. Once cell confluence was reached, the membranes were removed to initiate migration (\cref{fig:experiments}a). The experimental protocol and detailed analysis methodology can be found in the Methods. Briefly, monolayer boundaries were segmented from phase-contrast images at different time points (\cref{fig:experiments}b,c). Then, averaged $|y|$ displacements in the front and rear regions ($|y|$ in \cref{fig:experiments}d) were computed by dividing the enclosed area between the boundary at time $t$ and the initial boundary at $t=0$ h, by the width of the considered region ($x$ in \cref{fig:experiments}d). The evolution of a representative example is shown in \cref{fig:experiments}e. Finally, the averaged $|y|$ displacements were obtained by averaging over the same initial patterns and cell types. Measuring front and rear displacements in this way was much less sensitive to the finger-like shape fluctuations than a direct measure of the displacement of the center of mass, and allowed for a better comparison with simulations.

Anisotropic spreading of cell clusters, as described by Mode 3 in \cref{fig:modes}c, was consistently observed in patterned monolayers of MDCK (Madin-Darby canine kidney) and MCF-10A (human mammary epithelial) cells (\cref{fig:experiments}f, \cref{fig:experiments_supp_MDCKC3}, and \cref{fig:experiments_supp}). 

Theoretical fits from our data in \cref{fig:experiments}f show finite values of $L_c = 35 \pm 10 \;\mu$m, which are only consistent with an anisotropic spreading mode,  with a finite center-of-mass velocity. In this case, data were obtained from the large semicircle patterns for the MDCK cells, as they allow for optimal comparison with the simulations, both in terms of statistics and the well-defined shapes at early stages of the 2D monolayers. Moreover, the model predicts that semicircular shapes like these exhibit the largest motilities (see \cref{fig:evolutions}a). For our system size, we obtain of $V_{CM} = 4 \pm 1 \; \mu$m/h. Note that the limiting case defined as Mode 2 in \cref{fig:modes}b, also exhibiting a differential front-rear velocity, is ruled out by the unambiguous fit of a finite value $L_c$. This extends a few cells and is consistent with independent estimations from other experiments \cite{Blanch-Mercader2017b,Pérez-González2019}. For other shapes and sizes we did not reach significant statistics. However, we can use the fitted $L_c$ in \cref{fig:experiments}f to infer that a systematic differential front-rear velocity in the combined shapes and sizes is indicative of the anisotropic spreading mode (see this systematic trend in \cref{fig:experiments_supp}). A significant positive spreading displacement is observed for both cell lines during the first few hours of evolution, hence demonstrating the anisotropic spreading mode. The effect is somewhat smaller in the MCF-10A samples, which we attribute to their apparent higher surface tension, causing the monolayers to quickly revert to a circular form.

The choice of MDCK cells was appropriate for demonstrating Mode 3 for practical reasons and because of the good knowledge of their material parameters from previous studies \cite{Blanch-Mercader2017b}. Unfortunately, the low contractility of MDCK monolayers prevented a similar systematic study of Mode 4 (shown in \cref{fig:modes}d), placing the system far above the spreading transition for the sufficiently large clusters required by the continuum description to make sense. Nevertheless, carefully revisiting data from previous experiments by our group with other cell lines has allowed to identify examples of Mode 4 coherent migration\cite{Pallarès2023}. We thus provide five representative examples in a variety of situations where Mode 4 is relevant. In \cref{fig:experiments}g-k, we show unpatterned A431 cell clusters (human epidermoid carcinoma cells) where we identified instances of spontaneous migration associated with the rear-edge retraction whenever the cluster shape exhibited appropriate polar asymmetry. Moreover, in \cref{fig:experiments}i,j the motion is overcoming the durotactic migration caused by a stiffness gradient, as we predict in \cref{sec:spont_motility}. In \cref{fig:experiments}k the cluster moves perpendicularly to the stiffness gradient apparently guided by its asymmetry, in line with the model prediction. These unpatterned clusters share similar shapes with those in our controlled experiments, although typically smaller. They likely exhibit enhanced contractility, both due to the EGF treatment \cite{Chan2021, Iwabu2004} and their culturing in low adhesion plates, so that cell-cell adhesions are stronger than cell-matrix adhesions (see Methods). These observations are consistent with the coherent migration of Mode 4, as dictated by the asymmetric shapes in the presence of significant contractility. 

\section{Discussion and conclusion} \label{sec_discussion}

In this work, we have shown that confluent monolayer clusters are generically motile without external symmetry breaking. Clusters are capable of sensing their global shape as the only source of asymmetry and collectively migrate accordingly, thanks to the transmission of mechanical forces. For tissues with fast spreading (where traction forces dominate over cell-to-cell contractile forces), the effect is significant for small sizes and decreases over time as the cluster spreads. In contrast, if the total cluster area does not change significantly (when contractile forces are comparable to traction, or for low-compressibility tissues), the shape evolution yields a positive feedback that accelerates motion. The circular shape thus becomes morphologically unstable, so that, in the presence of noise, it will deform to acquire sustained motion through spontaneous symmetry breaking. The spontaneous motility of polar-asymmetric clusters is a direct consequence of the endogenous cue that polarizes epithelial cells at the cluster's edge so that they tend to move outward. A key point is the penetration of the polarizing cue in the form of a polarization boundary layer, that typically extends a few cells, due to the aligning interactions between cells. The penetration of the endogenous cue is sufficient to drive the coherent motion of all cells in the cluster if the global shape is asymmetric. We remark that the mechanism by which cells can sense the global shape is purely mechanical. It arises from the long-range transmission of hydrodynamic interactions, where active forces and cell-alignment forces at the polarized boundary layer are transmitted throughout the system, via the viscous stresses that encode cell-cell adhesion forces.

While it is unclear to what extent the spontaneous motion in an unstructured environment has a specific biological function, we emphasize that this endogenous effect will coexist and possibly compete with exogenous cues that might be present. We emphasize that in the appropriate ranges of parameters, the shape-induced velocities are relatively large. In particular, we have shown how shape-sensing motility interferes with a collective durotaxis response in realistic situations, enhancing it or even inducing a contrary motion. The latter clearly illustrates the collective nature of the effect here unveiled, implying that cells belonging to the cluster may be migrating in a different direction from that of isolated cells at the same location. 

In our theoretical framework, we have identified and explained the different physical scenarios that are possible in terms of the relevant parameters of a  minimal continuum model, which was previously tested for spreading monolayers. Our key prediction is that a finite velocity of the center of mass is generically expected for polar-asymmetric cluster shapes. We tested this prediction with a series of experiments under controlled conditions. Specifically we showed that, even in the presence of significant shape fluctuations, the statistics of cluster front and rear velocities provide, in the framework of our theoretical model, a precise measurement of the nematic screening length $L_c$ that quantifies the penetration of the polarization cue. This determination is instrumental to conclude on the motion of the center of mass of the cluster.

Overall, with the help of a theoretical framework, our results provide physical insights into the dynamics of epithelial monolayers and how cell interactions condition their collective behavior. These results contribute to our understanding of the physical mechanisms driving tissue organization and dynamics. Specifically, our work unveils a deep connection between shape and motion at the level of tissues, reminiscent of similar observations at the individual cell scale \cite{Verkhovsky1998, Blanch-Mercader2013}, with strong implications for the collective response of tissues to external cues in realistic situations. Finally, our results pave the way for future exploration of the interplay of endogenous and exogenous migration through the shape evolution of deformable cell clusters in natural or designed environments. Specifically, it also opens new avenues for the design and optimal control and manipulation of tissues for engineering applications.

\bigskip
\textbf{Acknowledgements.}  We thank all the members of our groups for their discussions and support. We thank Mònica Purciolas for technical assistance, and Meng Wang for his tips with the PDMS membranes. This paper was funded by the Generalitat de Catalunya (AGAUR SGR-2017-01602 to X.T., AGAUR SGR-2017-1061 to J.C., the CERCA Programme, and “ICREA Academia” award to J.C.); Spanish Ministry for Science and Innovation MICCINN/FEDER (PGC2018-099645-B-I00 to X.T., PID2019-108842GB-C21 to J.C., FPU19/05492 to I.P-J., PRE2020-092665 to J.T). European Research Council (Adv-883739 to X.T.); Fundació la Marató de TV3 (project 201903-30-31-32 to X.T.); European Commission (H2020-FETPROACT-01-2016-731957 to X.T.); the European Union’s Horizon 2020 research and innovation program (H2020 EpiFold to X.T.). La Caixa Foundation (LCF/PR/HR20/52400004 to X.T.); IBEC is recipient of a Severo Ochoa Award of Excellence from the MINECO.  

\smallskip
\textbf{Author contributions.} J.T., I.L. and J.C. conceived the project and developed the theory. I.P-J. performed experiments on patterned MDCK and MCF-10 cells and analyzed the data, and M.M. supervised experiments and analyzed the data. I.C.F. performed experiments on unpatterned A431 cells. All authors wrote the manuscript and revised the completed manuscript. X. T. supervised the experiments and J.C. supervised the project.

\smallskip
\textbf{Competing interests.} 
The authors declare no competing financial interests.

\smallskip
\textbf{Code availability.}
Analysis procedures and code implementing the model are available from the corresponding authors on reasonable request.

\smallskip
\textbf{Data availability.}
The data that support the findings of this study are available from the corresponding authors on reasonable request.

\smallskip
\textbf{Extended Data.}
Is available for this paper.

\smallskip
\textbf{Correspondence and requests for materials} should be addressed to J.C.

\bibliography{bibliography} 

\clearpage
\section*{METHODS}
\label{methods}
\setcounter{subsection}{0}
\renewcommand{\thesubsection}{\Alph{subsection}}

\subsection{Experimental Methods}

\textbf{Cell culture.}
Both MDCK (epithelial cells from the kidney tubule of an adult Cocker Spaniel dog) and MCF-10A cells (human epithelial cells from fibrocystic breast of an adult female), were cultured in Dulbecco’s Modified Eagle’s Medium containing high glucose and pyruvate (11995, Thermofisher) supplemented with 10$\%$ fetal bovine serum (FBS; Gibco), 1$\%$ penicillin and 1$\%$ streptomycin. Cells were maintained at 37$^\circ$C in a humidified atmosphere containing 5$\%$ CO$_2$.

\textbf{Microfabrication of PDMS membranes.}
Polydimethylsiloxane (PDMS) membranes were fabricated according to procedures described previously \cite{Wang2002, Poujade2007, Serra-Picamal2012, Sunyer2016, Uroz2018}. Briefly, SU8-50 masters containing our desired patterns (see layout and sketches in \cref{fig:experiments_patterns}) were raised using conventional photolithography. Uncured PDMS was spin-coated on the masters to a thickness lower than the height of the SU8 feature (approximately 100 $\mu$m), left overnight, and then cured for 1 h at 95$^\circ$C. PDMS was then peeled off from the master and kept in ethanol $95\%$ until use.

\textbf{Preparation of PAA gel substrates.} 
Polyacrylamide (PAA) gel preparation was adapted from previous protocols \cite{Yeung2005, Kandow2007}. Glass-bottom dishes were activated by using a 1:1:14 solution of acetic acid/bind-silane (M6514, Sigma)/ethanol $95\%$ for an hour (silanization). The dishes were washed twice with ethanol $95\%$ and dried by aspiration. Glass coverslips of 18 mm in diameter were treated with Repel Silane (General Electric, USA) for an hour, and then washed thoroughly, dipping three times and moving slowly into ethanol $95\%$, and three more times into Milli-Q water. They were then air-dried. A stock solution containing a concentration of 7.46$\%$ acrylamide, 0.044$\%$ bisacrylamide, 0.5$\%$ ammonium persulphate (APS) and 0.05$\%$ tetramethylethylenediamine (TEMED) is prepared to produce 5 kPa gels \cite{Tse2010}. TEMED must be added last since it triggers polymerization. A $22.5\;\mu$l drop of the solution, to have gels approximately $100\;\mu$m thick, was placed in the center of the glass-bottom dishes. The solution was then covered with the treated 18 mm coverslips to evenly distribute the gel and create a flat surface. After one hour, the polymerized PAA mixture was immersed in phosphate-buffered saline (PBS) for several minutes, and the coverslips were carefully removed with tweezers. The gels were rewashed in PBS and stored at 4$^\circ$C until use (at most three weeks later).

\textbf{PAA gel functionalization with ECM protein.} 
PAA gels were incubated with a solution of 2 mg·ml$^{-1}$ Sulpho-SANPAH in Milli-Q water under UV for 7.5 min (365 nm wavelength, at a distance of 5 cm). Excess Sulfo-SANPAH was removed with two consecutive washes with HEPES pH $=$ 7.0 buffer of 2.5 min each, and a quick last one with PBS. After air-drying for 5 min, the gels were incubated overnight at 4 $^\circ$C with 100 $\mu$l of rat tail type I collagen solution (0.1 mg·ml$^{-1}$, Millipore). The gels were then UV-sterilized for 20 minutes before cell seeding.

\textbf{Cell patterning on PAA gels.} 
The PDMS membranes were air-dried and then passivated by incubation with a solution of pluronic acid F127 2$\%$ in PBS, shaking for one hour. After incubation, they were washed twice in PBS, and well dried for 20 min. Meanwhile, the collagen-coated PAA gels were washed twice with PBS and completely dried, first by aspiration and then air-dried, for no more than 8 min. Passivated PDMS membranes were carefully placed in the center of the gels, and then $400\:\mu$l of medium with cells, with a density $\sim 750\cdot 10^3$ cells/ml, was placed on top of the membranes, covering them all. After 30 min, the unattached cells were washed away and fresh medium was added. Cells attached to the gel only at the openings of the membranes. Once they reached confluence (typically between 10 and 20 h), the membrane was peeled off, allowing the cells to migrate freely over the surrounding space. A schematic of the steps is found in \cref{fig:experiments}a.

\textbf{Time lapse microscopy.}
As soon as the confinement was released by removing the PDMS membranes, the samples were transferred to the microscope, and time-lapse imaging typically began about an hour after the release. Multidimensional acquisition routines were performed on automated inverted microscopes (Nikon Eclipse Ti) equipped with thermal, CO$_2$, and humidity control, operating with the MetaMorph software. The image acquisition interval was 15 min, and a typical experiment was run for at least 14 h. Images were acquired using a 10X 0.3 NA objective (Nikon Plan Fluor 10X/0.30 Ph1 DLL WD 16), and an automated stage was employed to capture multiple positions. Phase-contrast images of the migration of the monolayers were saved for subsequent analysis. 

\textbf{Analysis of the phase-contrast images.}
Custom-made MATLAB scripts combined with registration plug-ins from Fiji software were used to process the phase-contrast images. Some of the samples had experimental issues, like peeling of the collagen layer after the removal of the PDMS membrane, too crowded or not confluent monolayers due to differences in sizes, or too many multicellular protrusions in the initial stages of the migration. These samples were excluded from the analysis since they were not suited to compare with the theoretical results. Rotation patterns (E in \cref{fig:experiments_patterns}) were also excluded because no rotation was observed. The numbers of selected samples for the analysis, corresponding to each pattern design and cell type can be found in \cref{fig:experiments_patterns}.

At every time point, images from the selected samples were first registered (Fiji plug-ins ``StackReg'' with the translation mode or `Correct3Ddrift'' in the $xy$ plane and ``DescriptorBased'' with the Rigid 2D mode), and then rotated and translated using custom-made MATLAB codes, to have centered patterns in the images, with their axis of mirror symmetry aligned along the vertical $y$ axis. Then, the islands were semi-automatically segmented (using the ``EGT$\_$Segmentation'' function from MATLAB, developed at the National Institute of Standards and Technology, and manually corrected with Fiji for all those that presented errors).

To average out details, the $|y|$ displacement was computed by dividing the enclosed area between the boundary at time $t$ and the initial boundary at $t=0$ h, by the width of the considered region (\cref{fig:experiments}d). This region encompasses $\sim10$ cells in width around the top and bottom points (those vertically aligned with the center of the cluster). Evolutions of the averaged $\abs{y}$ displacements (\cref{fig:experiments}e) were then averaged within each pattern and cell type to calculate the averaged $\abs{y}$ displacements (\cref{fig:experiments}f and \cref{fig:experiments_supp_MDCKC3} for the averaged and individual realizations of the C3 MDCK samples, and \cref{fig:experiments_supp} for the combined patterns and sizes for the different cell lines). 

\textbf{Unpatterned A431 clusters.}
Data presented here for unpatterned A431 clusters (\cref{fig:experiments}g-k) is from experiments published in Ref. \cite{Pallarès2023}, where more details can be found in the Methods section of that reference. Briefly, A431 cells (human epidermis cells from an epidermoid carcinoma of an adult female) were cultured in the same medium as MDCK and MCF-10A cells. Cell clusters were obtained by seeding $5\cdot 10^3$ cells per well in Corning Costar ultra-low attachment multiple well plates (CLS3474-24EA) for 24 h in starvation media (Dulbecco’s Modified Eagle’s Medium containing high glucose and pyruvate supplemented with 1$\%$ FBS, 1$\%$ penicillin and 1$\%$ streptomycin). They were then mechanically disaggregated into smaller clusters exhibiting heterogeneous sizes by pipetting up and down. Cellular debris was discarded, and the clusters were resuspended in media containing 5 $\mu$M RHO/ROCK pathway inhibitor (Y-27362). They were then seeded onto fibronectin-coated PAA gels in a total volume of 50 $\mu$l. After 45 minutes, 1 ml of starvation media was added to prevent the gels from drying out. Immediately before imaging the clusters 2 hours later, starvation media containing hEGF was added to the plates to achieve a final concentration of 1 ng/mL, enhancing cellular contractility \cite{Chan2021, Iwabu2004}.

For uniform stiffness PAA gels of $30$ kPa, a 1 ml gel premix solution containing 125 $\mu$l of 2$\%$ bis-acrylamide and 244 $\mu$l of 40$\%$ acrylamide, 7.5 $\mu$l irgacure 5$\%$ w/v (BASF), 6 $\mu$l acrylic acid (147230, Sigma-Aldrich), 84 $\mu$l 1 M NaOH and 10 $\mu$l 500-nm-diameter yellow–green fluorospheres was prepared (F8813, Thermofisher). A drop of 16 $\mu$l of gel premix was added to the center of the previously silanized glass-bottomed dish and a 18-mm-diameter glass coverslip treated with Repel Silane was placed on top. It was then placed under UV light for 10 min to allow gel polymerization. Gradients of stiffness in the PAA gels were obtained with a gel premix with 15$\%$ acrylamide, 1$\%$ biacrylamide, 0.75 mg$\cdot$ml$^{-1}$ irgacure, 0.60$\%$ acrylic acid, 100 mM NaOH and a dilution of 1:100 from 500-nm-diameter fluorescent beads. A 25 $\mu$l drop was added to the center of previously silanized glass-bottomed dishes and covered with an 18-mm-diameter glass coverslip treated with Repel Silane. The stiffness gradients were produced by making use of an opaque sliding mask during UV-triggered gel polymerization for 5 min \cite{Sunyer2012, Sunyer2016}. In both cases, after gel polymerization 10x PBS was added and the top coverslips were separated from the gels with round-tipped tweezers.

Finally, PAA gels were functionalized using carbodiimide reactions. They were incubated with 100 mM EDC and 200 mM NHS in 20 mM HEPES pH$=$7.0 buffer at 37$^\circ$C for 20 min. Next, they were quickly washed twice with PBS and incubated at 37$^\circ$C with a dilution of 0.1 mg$\cdot$ml$^{-1}$ fibronectin in PBS for 45 min. Finally, gels were washed twice with PBS and incubated with 1 M Tris pH$=$8.0 for 30 min at room temperature, followed by two PBS washes.

\subsection{Numerical Scheme}\label{sec:numerics}
\renewcommand{\thesubsubsection}{\Roman{subsubsection}}

Here we describe our method for numerically integrating \cref{eq:quasi_static_solution} and \cref{eq:velocity} and evolving the domain $\Omega(t)$. 

Before proceeding with the numerical integration we adimensionalize the equations. We take $L = \sqrt{A/\pi}$ as the characteristic system size, the characteristic time scale of a flat front expansion without contractility, $\tau = \eta/(L_c\,\zeta_\text{i})$, as the time scale, and $L_c\,\zeta_\text{i}$ as the characteristic stress. Then, $\mathbf{p}=\mathbf{p}'$ and $\mathbf{v} = (L\cdot L_c \cdot \zeta_\text{i}/\eta)\,\mathbf{v}'$, so the equations of the simplified model; \cref{eq:quasi_static_solution}, \cref{eq:constitutive}, and \cref{eq:external} (with $\nabla \Pi \approx 0$) become

\begin{align}
    \nabla'^2\mathbf{p}' - \left(\frac{L}{L_c}\right)^2\mathbf{p}' &= 0 \label{eq:adim_polarity}\\ 
    \nabla'\cdot\mathbf{\sigma}' - \left(\frac{L}{\lambda}\right)^2\mathbf{v}'+ \frac{L}{L_c}\mathbf{p}' &= 0  \label{eq:adim_velocity}\\
    \left[(\nabla' \mathbf{v}' + \nabla' {\mathbf{v}'}^T) + \frac{\ell_a}{L_c} \;\mathbf{p} \mathbf{p}\right] &= \mathbf{\sigma}' \label{eq:adim_stress},
\end{align}

where $\lambda=\sqrt{\eta/\xi}$ is the hydrodynamic screening length and $\ell_a = |\zeta|/\zeta_\text{i}$ is the active length.

With these equations, the model parameters become the three characteristic scales ($L$, $\tau$ and $L_c\,\zeta_\text{i}$) and three ratios: $L_c/L$, $\lambda/L$ and $\ell_a/L$. From now on, we will omit the primes of the adimensional fields for clarity.

We solve the model in three steps: given the $n$-th timestep, where $\Omega^n$ is the cluster domain, we start by computing the polarity, which characterizes the active forces in \cref{eq:adim_velocity}, then we integrate it to find the velocity, and finally we evolve the domain using the kinematic condition $V_n = \mathbf{v}\cdot \hat{\mathbf{n}}$. The domain $\Omega^n$ is discretized into an unstructured mesh, $\mathcal{T}_h^n$, with an average separation $h$ between the vertices of the triangle. Polarity and velocity fields are integrated with the finite element method (FEM), using the open source package FreeFem++\cite{Hecht2012}, and then analysed and visualized using a post-processing code written in Mathematica (TM).

\textbf{Solving the polarity field.} 
To numerically integrate \cref{eq:adim_polarity} with the boundary condition $\mathbf{p}|_{\partial\Omega} = \mathbf{\hat{n}}$, which imposes an inhomogeneous Dirichlet problem, we start by converting it into a modified homogeneous Dirichlet problem for $\mathbf{p}_H = \mathbf{p}-\mathbf{p}_D$:

\begin{align}
    &\nabla^2 \mathbf{p}_H - \frac{L^2}{L_c^2}\,\mathbf{p}_H = -\nabla^2 \mathbf{p}_D + \frac{L^2}{L_c^2}\,\mathbf{p}_D & \text{in } \Omega(t),\label{eq:homogeneous_pol}\\
    &\mathbf{p}_H = 0 & \text{in } \partial\Omega(t),
\end{align}

where $p_{Dx},\,p_{Dy} \in H^{1}(\Omega)$ such that $\mathbf{p}_D = \mathbf{\hat{n}}$ in the boundary, $\partial\Omega(t)$. 
To proceed with the FEM, we write the weak form of \cref{eq:homogeneous_pol},

\begin{align}
    &\left(\frac{L}{L_c}\right)^2\,\int_\Omega \mathbf{p}_H\cdot \mathbf{q} \;\mathrm{d}S + \int_\Omega \nabla\mathbf{p}_H\cdot \nabla\mathbf{q} \;\mathrm{d}S = \nonumber\\
    & \qquad -\left(\frac{L}{L_c}\right)^2\,\int_\Omega \mathbf{p}_D\cdot \mathbf{q} \;\mathrm{d}S - \int_\Omega \nabla\mathbf{p}_D\cdot \nabla\mathbf{q} \;\mathrm{d}S,
    \label{eq:weak_pol}
\end{align}

where we multiplied \cref{eq:homogeneous_pol} by the test function $\mathbf{q}$, integrated by parts over the domain, and apply the homogeneous boundary condition for $\mathbf{p}_H$. Then, the problem consists of finding a solution $p_{Hx},\,p_{Hy} \in H_0^1(\Omega)$ for any test functions $q_x,\,q_y \in H_0^1(\Omega)$. The solution for the polarity field is then given by $\mathbf{p} = \mathbf{p}_H + \mathbf{p}_D$.

The LHS terms are bilinear (implicit), and those on the RHS are linear (explicit). In our code, we represent $p_{Hx},\,p_{Hy}$ and $q_x,\,q_y$ by the basis functions spanning the continuous \textbf{P2} finite-element space, i.e., quadratic polynomials defined piece-wise on each element $K \in \mathcal{T}_h^n$. We chose to use \textbf{P2} elements, which have a smooth gradient, to properly resolve the active contractile stress at \cref{eq:adim_velocity}. The matrix defined by the LHS is inverted using the default \textbf{sparsesolver} in FreeFem++\cite{Hecht2012}.

\textbf{Solving momentum balance.}
In order to solve the velocity we numerically integrate \cref{eq:adim_velocity}, the momentum balance, together with stress-free boundary conditions, $\mathbf{\sigma}\colon \mathbf{\hat{n}} \mathbf{\hat{n}} =0$. Deriving the weak formulation of this problem is straightforward, as $\mathbf{\sigma}\cdot \mathbf{\hat{n}}$ appears naturally after integrating by parts. Multiplying \cref{eq:adim_velocity} by the test function $\mathbf{u}$, integrating by parts and substituting \cref{eq:adim_stress},

\begin{align}
    &\int_\Omega (\nabla\mathbf{v}+\nabla\mathbf{v}^T)\,\colon\,\nabla\mathbf{u} \;\mathrm{d}S + \left(\frac{L}{\lambda}\right)^2\int_\Omega \mathbf{v}\cdot\mathbf{u} \;\mathrm{d}S = \nonumber \\
    & \qquad \frac{L}{L_c}\int_\Omega \mathbf{p}\cdot\mathbf{u} \;\mathrm{d}S - \frac{\ell_a}{L_c}\int_\Omega \mathbf{p}\mathbf{p}\,\colon\,\nabla\mathbf{u} \;\mathrm{d}S,
    \label{eq:weak_vel}
\end{align}

where we have rearranged the terms so that \cref{eq:weak_vel} has a bilinear (implicit) LHS and a linear (explicit) RHS. To conclude, the problem consists of finding $v_x,\, v_y \in H^1(\Omega)$ such that \cref{eq:weak_vel} holds for any test function $u_x,\, u_y \in H^1(\Omega)$. In this case we opt to approximate $v_x,\,v_y$ and $u_x,\,u_y$ in the code by the basis functions spanning the continuous \textbf{P1b} finite-element space, i.e.,linear polynomials enriched with a \textit{bubble} function (a cubic polynomial defined as the product of the barycentric coordinates in the element $K$ and vanishing on its faces) defined piecewise for each element $K \in \mathcal{T}_h^n$. Again, the LHS matrix is inverted using \textbf{sparsesolver} in FreeFem++\cite{Hecht2012}.

\textbf{Remeshing and domain-evolution.}
After integrating the polarity and velocity, we evolve the domain $\Omega^n$ to $\Omega^{n+1}$ by moving each vertex of every element $K\in\mathcal{T}_h^n$ by $\mathbf{v}\,dt$, keeping $\mathbf{v}\,dt < h$. Then, we take the new boundary, $\partial\Omega^{n+1}$, remesh it using akima splines, provided by the GLS library\cite{galassi2018}, and use it to compute a new mesh $\mathcal{T}_h^{n+1}$. With this procedure, we guarantee that the average spacing between vertices remains close to $h$ upon evolving the domain.

\textbf{Additional notes.}
\vspace{-0.2cm}
\begin{itemize}
    \item As commented in the main text, the polarity field in the present model decays inwards form the boundary with a characteristic length of $L_c/L$ (in adimensionalized coordinates), and thus most of the model dynamics occur in a boundary layer of this depth. Consequently, and in order to optimize the computation time and memory, we opt to define adaptive (nonuniform) meshes, designed to have a finer density of vertices along the boundary and a coarser one in the bulk, inwards form the boundary layer.
    
    \item Note that the code considers a polygonal approximation of the domain boundaries, so the normal vector to the boundary is well defined in the edges, but not in the vertices. The velocity field is very sensible to systematic shifts in the definition of the normal vectors and thus we redefine them in terms of the normals of the neighbouring edges. Consider a vertex $i$ with a previous edge of length $\ell_{-1}$ and normal $\mathbf{\hat{n}}_{-1}$ and a posterior edge with $\ell_1$ and $\mathbf{\hat{n}}_1$, then we define the normal at the vertex as
    
    \begin{equation}
        \mathbf{n}_i=\frac{1}{1/\ell_1 + 1/\ell_{-1}}\left(\frac{\mathbf{\hat{n}}_1}{\ell_1} +\frac{\mathbf{\hat{n}}_{-1}}{\ell_{-1}}\right)
    \end{equation}

    and normalize it. This definition ensures that the normals at the vertices will be biased towards the normals at the shorter edges. 
\end{itemize}

\begin{table}[htb!] 
    \centering
    \begin{tabular}{l|c|c}
    Simulations & $R_0$ ($\mu$m) & $\theta_c$ (rad) \\
    \noalign{\smallskip}\hline\noalign{\smallskip}
    \cref{fig:evolutions} \textbf{a} & 200 & $\pi$/6 - 4$\pi$/3 \\
    \cref{fig:evolutions} \textbf{d},\textbf{e},\textbf{h},\textbf{i}$^*$ & 166 & 2$\pi$/3 \\
    \cref{fig:evolutions} \textbf{f},\textbf{g},\textbf{j},\textbf{k}$^*$ & 145 & $\pi$/6 \\
    \cref{fig:evolutions} \textbf{l} & 261 & $\pi$ \\
    \cref{fig:evolutions} \textbf{m} & 261 & 2$\pi$/3 \\
    \cref{fig:modes} \textbf{e}-\textbf{i} & 200 & $\pi$ \\
    \cref{fig:friction} \textbf{a}-\textbf{c} & 200 & $\pi$ \\
    \cref{fig:experiments} \textbf{f}$^{**}$ & 282.843 & $\pi$ \\
    \end{tabular}
    \caption{\textbf{Shape parameters for the different simulations in the figures.} $\mathbf{^{*}}$The radius were chosen so that the simulations start close to the spreading transition. $\mathbf{^{**}}$This is the radius that gives an effective size $L=\sqrt{A/\pi}$ of $200 \,\mu\text{m}$ for a semicircle.} 
    \label{tab:params_shape}
\end{table}

\vspace{-0.2cm}
\textbf{Numerical parameters.}
The non-chiral simulations have an initial domain constructed as a circle of radius $R_0$ ($R_0/L$ in adimensionalized units) with a straight cut of angular length $\theta_c$ and rounded corners with a radius of a 10\% of the base circle radius. The chiral cluster at \cref{fig:evolutions}l has been constructed as the non-chiral shapes, but reversing the left half of the shape in the $y$-direction. See the exact values of $R_0$ and $\theta_c$ for the different simulations at \cref{tab:params_shape}.

The outer boundary of the shapes has a vertex density given by $\rho_{outer} = 45$ vertices per adimensionalized unit of the boundary length for all the simulations except for the ones with $L_c < 5 \,\mu\text{m}$ at \cref{fig:friction}, that have $\rho_{outer} = 600$, so $L_c/L\gtrsim h$. The inner boundary of the boundary layer has a density $\rho_{inner} = 15$, while for the simulations with small $L_c$ at \cref{fig:friction}, that have $\rho_{inner} = 100$.

The simulations showing temporal evolutions have a timestep $dt = 5\cdot10^{-4}$ in adimensionalized time units.

\textbf{Verification.}
We verified our computational solver for different geometries and parameter values. We tested a rectangular geometry with periodic boundary conditions along the $x$-direction against the analytical solutions for an infinite stripe\cite{Alert2019a}, in the whole range of friction values (from $\lambda < L$ to $\lambda \gg L$), and the results of a circle with large screening length (wetting limit) with the analytical solutions for spreading and retracting clusters\cite{Pérez-González2019}.

\vfill

\clearpage
\onecolumngrid 
\section*{FIGURES}

\begin{figure}[htb!]
    \centering
    \includegraphics[width=.5\columnwidth]{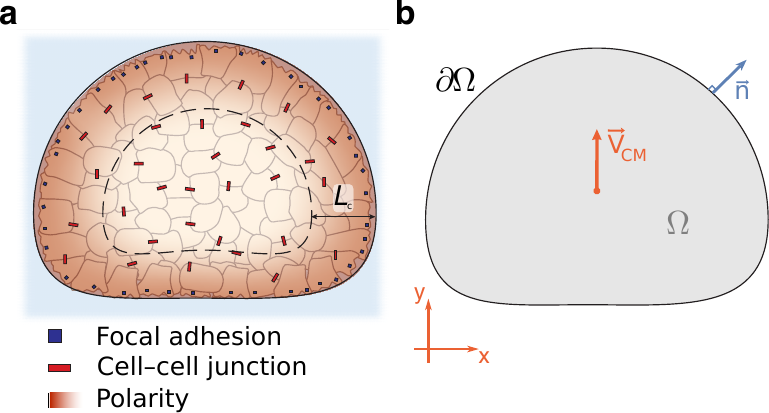}
    \caption{\textbf{Schematic representation of a cell cluster monolayer and its geometry} \textbf{a,} Sketch of a cell cluster monolayer with the biological cues that generate the model forces, namely focal adhesions and cell-cell junctions. \textbf{b,} Simplified geometry of a cell cluster that shows the position and velocity of the center of mass, the cluster domain ($\Omega$) and its boundary ($\partial\Omega$) and a boundary normal, which acts as a boundary condition for the polarity field (homeotopic anchoring).}
    \label{fig:sketch}
\end{figure}

\clearpage

\clearpage

\clearpage

\clearpage
\section*{SUPPLEMENTARY FIGURES}
\renewcommand{\thesection}{\Roman{section}}
\renewcommand{\thesubsection}{\Roman{subsection}}
\renewcommand{\thesubsubsection}{\Roman{subsection}.\arabic{subsubsection}}
\renewcommand{\thetable}{S\Roman{table}} 
\renewcommand{\theHtable}{S\Roman{table}} 
\renewcommand{\theequation}{S\arabic{equation}}
\renewcommand{\figurename}{Fig.}
\renewcommand{\thefigure}{S\arabic{figure}}  
\renewcommand{\theHfigure}{S\arabic{figure}} 
\setcounter{table}{0}
\setcounter{equation}{0}
\setcounter{figure}{0}
\setcounter{section}{0}
\setcounter{subsection}{0}

\begin{figure}[htb!]
    \centering
    \includegraphics[width=\columnwidth]{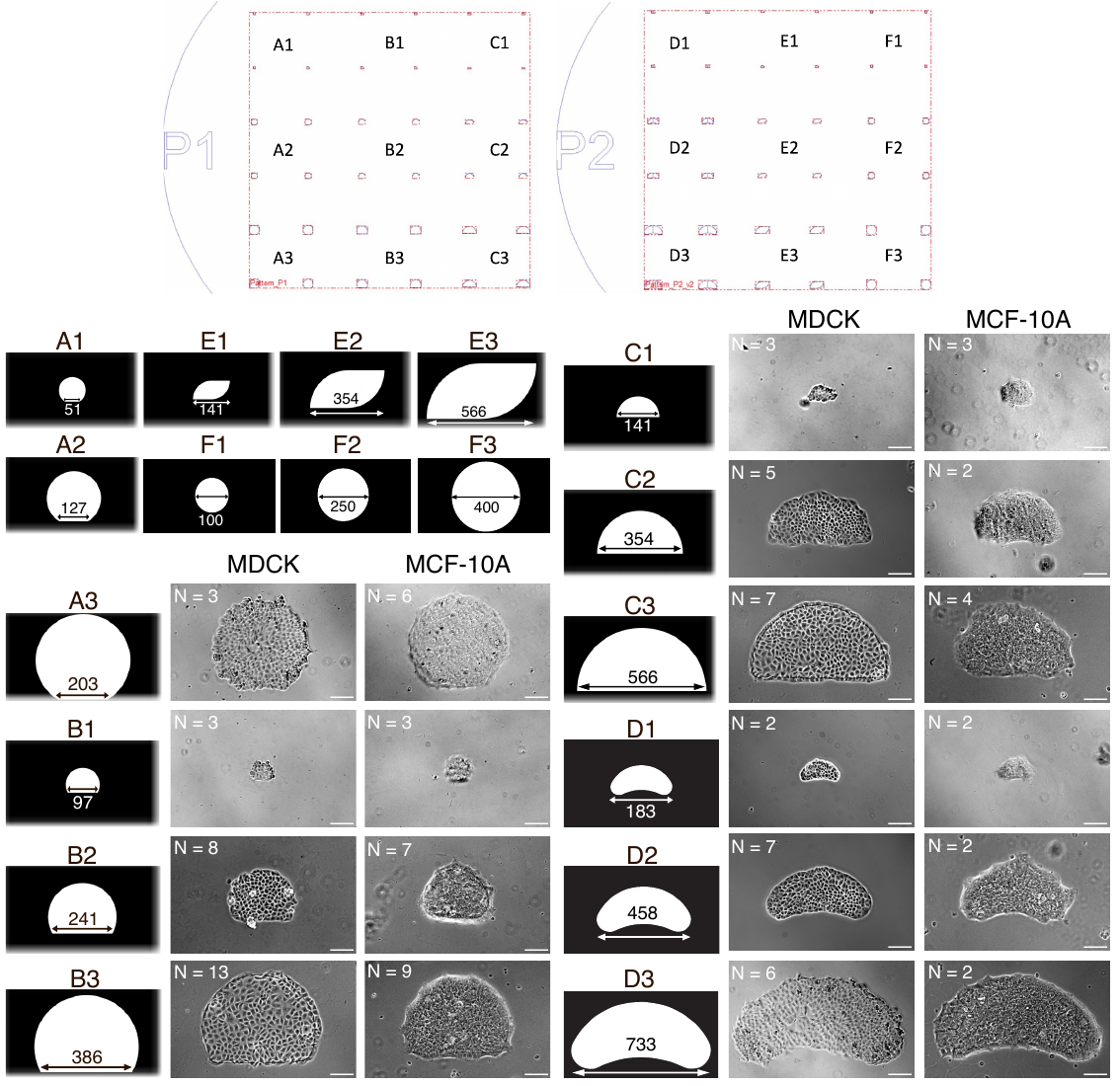}
    \caption{\textbf{Experimental patterns.} Layout of the patterns in the PDMS membranes (P1 and P2), which were used indistinctly for the experiments. Below, binaries of the patterns and sizes (in $\mu$m), being the effective radii (radius of a circle with the same area) $L = 50,125,200$ $\mu$m for 1,2 and 3 sizes respectively. For each pattern and cell line, we show a representative phase-contrast image of the initial time point (scale bars, $100\:\mu$m), being $N$ the number of analyzed samples (in total, $N_{\text{MDCK}}=57$ and $N_{\text{MCF-10A}}=40$). No samples were selected for the A1 and A2 patterns, and E and F were not analyzed since they were not suited to compare with the theory predictions.}
    \label{fig:experiments_patterns}  
\end{figure}

\begin{figure}[htb!]
  \centering
  \includegraphics[width=0.8\columnwidth]{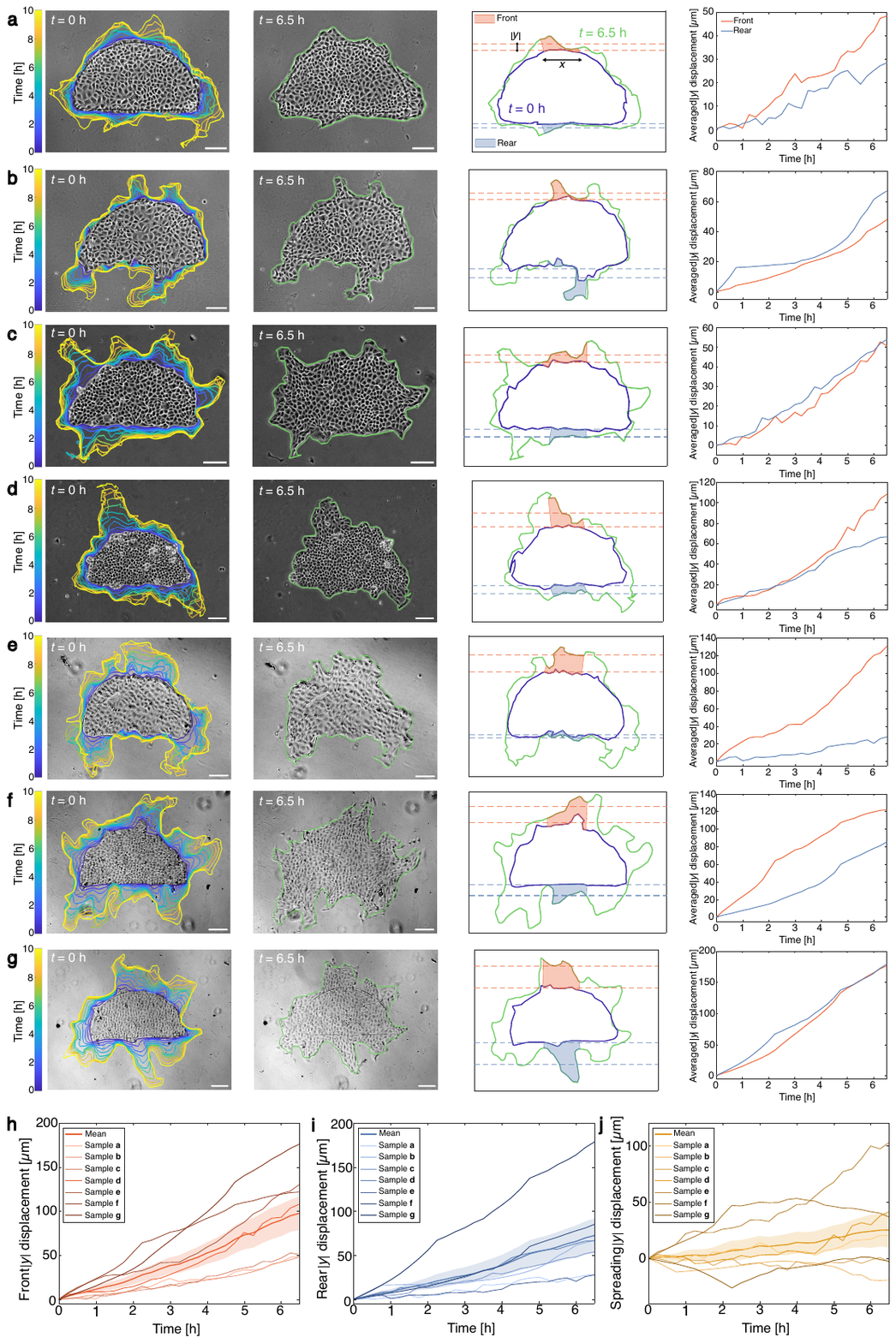} 
  \caption{\textbf{Selected C3 MDCK samples.} \textbf{a-g,} Phase-contrast images at $t=0$ h (first column, where each line is every hour) and $t=6.5$ h (second column), segmented masks and enclosed areas for the computation (third column), and temporal evolutions of the front (red) and rear (blue) averaged $\abs{y}$ displacements (fourth column, note that $y$-axis scales are different for the sake of visualization). Scale bars, $100\:\mu$m. \textbf{h-j,} Altogether, the averaged displacements of the individual samples are plotted for the front (\textbf{h}), rear (\textbf{i}) and spreading (\textbf{j}) displacements. Thicker lines are experimental means, and the shaded area is the Standard Error of the Mean (SEM). These data combined in one single plot give \cref{fig:experiments}f.}
  \label{fig:experiments_supp_MDCKC3}
\end{figure}

\begin{figure}[htb!]
  \centering
  \includegraphics[width=0.8\columnwidth]{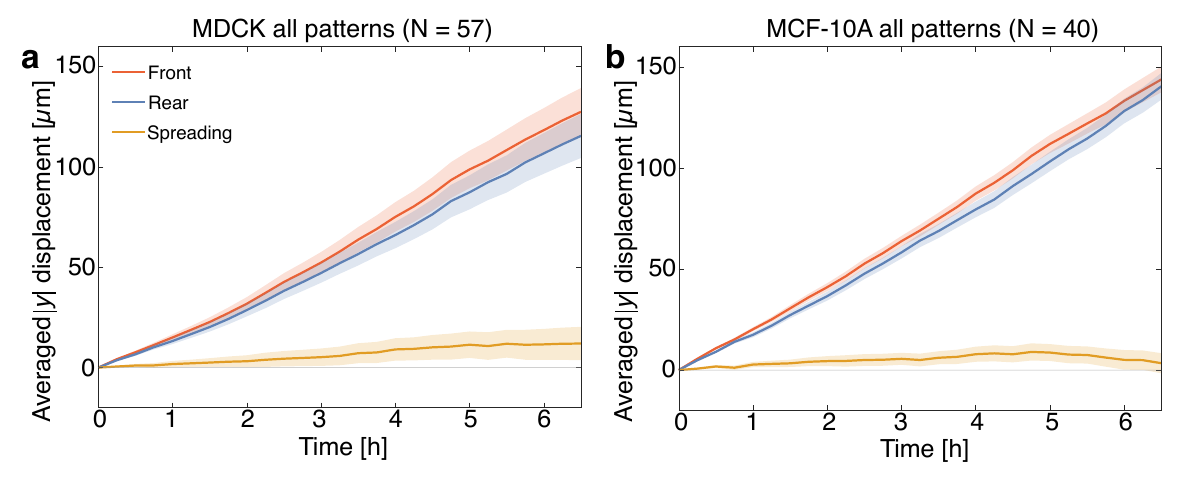} 
  \caption{\textbf{Averaged displacements for the combined patterns.} Front (red), rear (blue), and spreading (yellow) averaged displacements, when all the patterns and sizes are combined, for MDCK (\textbf{a}) and MCF-10A cell lines (\textbf{b}). The statistics of the combined patterns and sizes can be found in \cref{fig:experiments_patterns}.}
  \label{fig:experiments_supp}
\end{figure}

\clearpage
\section*{SUPPLEMENTARY MATERIAL}
\subsection{Kinematics of a cell monolayer as a 2D compressible flow}
The cell monolayer is a quasi-2D system, with a small thickness $h(\mathbf{r},t)$ in the third dimension. We model it as a 2D system, where the in-plane velocity $\mathbf{v}$ is defined by averaging over the third dimension. We may assume that the 3D fluid has a constant density $\rho$ and a mass growth rate density $\sigma_V$. Then, the surface mass density is $\rho h$ and the mass balance equation for the 2D averaged fluid reads

\begin{equation}
    \frac{\partial h}{\partial t}
    + \mathbf{\nabla}  \cdot (h \mathbf{v})
    = \sigma_S h,
    \label{eq:massbalance}
\end{equation}

where $\sigma_S=\sigma_V/\rho$. Consequently, the effective 2D fluid is generically compressible, i.e. $\mathbf{\nabla}\cdot \mathbf{v} \neq 0$, and the thickness of the monolayer can be inferred from the knowledge of $\mathbf{v}$ using \cref{eq:massbalance}. The basic observable that we address here is the geometric centroid of the 2D domain occupied by the tissue, $\Omega(t)$. The centroid is the aerial center of mass, as opposed to the physical center of mass. The former is the most suitable quantity to monitor the shape evolution and displacement of the domain $\Omega(t)$, since the information about $h(\mathbf{r},t)$, may not be available, in particular when comparing with experiments. In addition, the physical center-of-mass would account for mass rearrangements, including those that may not contribute to changes in the shape and location of $\Omega(t)$, which are not our focus of attention. In the following subsection we derive the velocity of the center of mass of $\Omega(t)$ defined as the geometric centroid.

\subsection{Harmonic moment expansion and derivation of the center-of-mass velocity} \label{SM:Vcm_derivation}
\titleformat{\paragraph}{\small\filcenter\itshape}{\theparagraph}{1em}{}
\titlespacing*{\paragraph}{0pt}{3.25ex plus 1ex minus .2ex}{1.5ex plus .2ex}
\setcounter{secnumdepth}{2}
\setcounter{table}{0}
\setcounter{equation}{0}
\setcounter{figure}{0}
\setcounter{section}{0}
\setcounter{subsection}{0}

The harmonic moment expansion of a free-boundary problem is a technique that allows to characterize the temporal evolution of an interface in terms of its global geometric properties. It has been  particularly useful in Laplacian free-boundary problems as in \cite{Cummings1999,Miranda2010,Blanch-Mercader2013}, but it is general and can be exploited also in our problem.

The temporal evolution of a free interface $\partial \Omega$ can be described by the harmonic moments of the region it encompasses, $\Omega$. We define the complex $k$-th moment as

\begin{equation}
    M_k = \int_\Omega z^k \,\mathrm{d}a,
    \label{complex_moment}
\end{equation}

with $z=x+iy$. Then, its temporal derivative takes the form \cite{Cummings1999,Miranda2010,Blanch-Mercader2013}

\begin{equation}
    \frac{\mathrm{d}M_k}{\mathrm{d}t} = \frac{\mathrm{d}}{\mathrm{d}t} \int_\Omega z^k \,\mathrm{d}a = \int_{\partial\Omega} z^k V_n\mathrm{d}l = \int_{\partial\Omega} z^k (\mathbf{v}\cdot \hat{\mathbf{n}})\,\mathrm{d}l,
\end{equation}

where in the last equality we have used the kinematic condition, $V_n = \mathbf{v}\cdot \hat{\mathbf{n}}$. Next, applying the divergence theorem and expanding it,

\begin{equation}
    \int_{\partial\Omega} z^k (\mathbf{v}\cdot \hat{\mathbf{n}})\,\mathrm{d}l = \int_\Omega \nabla \cdot (z^k \mathbf{v}) \,\mathrm{d}a =  \int_\Omega \left[ \mathbf{v}\cdot \nabla z^k + z^k (\nabla \cdot \mathbf{v})\right] \,\mathrm{d}a.
\end{equation}

Now, let's take a moment to analyze the temporal evolution of the zeroth moment;

\begin{equation}
    \frac{\mathrm{d}M_0}{\mathrm{d}t} = \int_\Omega \nabla \cdot \mathbf{v}\,\mathrm{d}a.
\end{equation}

It is easy to see from \cref{complex_moment} that the zeroth moment corresponds to the cluster area. Thus, the equation above generically characterizes cluster spreading (retraction), when ${\mathrm{d}M_0}/{\mathrm{d}t} > 0$ ($<0$).

Taking into account that the area does not remain constant and defining the $k$-th moment with respect to the center of mass, $M_k^{CM} = M_k /M_0$;

\begin{equation}
    \frac{\mathrm{d}M_k}{\mathrm{d}t} = \frac{\mathrm{d}}{\mathrm{d}t}\left(M_0M_k^{CM} \right) = \frac{\mathrm{d}M_0}{\mathrm{d}t}M_k^{CM} + M_0\frac{\mathrm{d}M_k^{CM}}{\mathrm{d}t}.
\end{equation}

Consequently, for $k\geq 1$,

\begin{align}
    \frac{\mathrm{d}M_k^{CM}}{\mathrm{d}t}
    &= \frac{1}{M_0}\left( \frac{\mathrm{d}M_k}{\mathrm{d}t} - \frac{1}{M_0}\frac{\mathrm{d}M_0}{\mathrm{d}t}M_k \right) \nonumber \\ 
    &= \frac{1}{A(t)}\left[ \int_\Omega \mathbf{v}\cdot \nabla z^k \,\mathrm{d}a + \int_\Omega z^k (\nabla \cdot \mathbf{v}) \,\mathrm{d}a -\frac{1}{A(t)}\int_\Omega (\nabla\cdot\mathbf{v})\,\mathrm{d}a \int_\Omega z^k\,\mathrm{d}a \right] \nonumber \\
    &= \frac{1}{A(t)}\left[ \int_\Omega \mathbf{v}\cdot \nabla z^k \,\mathrm{d}a + \int_\Omega \left(z^k-M_k^{CM}\right) \nabla \cdot \mathbf{v} \,\mathrm{d}a \right],
    \label{general MkCM}
\end{align}

Finally, the geometric velocity of the center of mass (the centroid) corresponds to the temporal derivative of the first harmonic moment of the center of mass. Substituting $k=1$ on \cref{general MkCM} we obtain \cref{eq:kinematics}, as shown in the main text.
\end{document}